\documentclass[11pt,a4paper]{article}
\usepackage[a4paper,width=170mm,top=25mm,bottom=25mm]{geometry}
\usepackage{amsmath,amssymb}
\usepackage{mathtools}
\usepackage[affil-it]{authblk}
\usepackage[english]{babel}
\usepackage{graphicx}
\usepackage{rotating}
\usepackage{adjustbox}
\usepackage{setspace}
\usepackage{verbatim}
\usepackage{caption}
\usepackage{wrapfig}
\usepackage{color}
\usepackage{subcaption}
\usepackage{booktabs}
\usepackage{bm}
\usepackage[toc,page]{appendix}
\usepackage{csquotes}
\usepackage[breaklinks=true,colorlinks=true,linkcolor=blue,urlcolor=blue,citecolor=blue]{hyperref}
\usepackage[
  backend=bibtex,
  style=numeric-comp,
  sorting=none,
  sortcites=true,
  maxbibnames=1,
  minbibnames=1,
  giveninits=true
]{biblatex}
\usepackage{orcidlink}
\usepackage{placeins}

\usepackage[normalem]{ulem}

\title{First results from a borehole network for the seismic characterization of the Einstein Telescope Euregio Meuse-Rhine candidate site}

\author[1]{Luca D'Onofrio\,\orcidlink{0000-0001-9546-5959}\thanks{Corresponding author: \texttt{l.donofrio@nikhef.nl}}}
\author[2]{Soumen Koley\,\orcidlink{0000-0002-5793-6665}}
\author[1]{Pablo Barneo,\orcidlink{https://orcid.org/0000-0002-8883-7280}}
\author[1]{Stan Bentvelsen\,\orcidlink{0000-0003-3280-0953}}
\author[1,5]{Stefan Danilishin\,\orcidlink{0000-0001-7758-7493}}
\author[3]{Geert Degrande\,\orcidlink{0000-0002-1763-4144}}
\author[3]{Stijn Fran\c{c}ois\,\orcidlink{0000-0002-3877-9748}}
\author[6]{S. Kadmiel\,\orcidlink{0000-0003-2215-6164}}
\author[1]{M. Kiehn}
\author[2]{Hadrien Michel\,\orcidlink{0000-0003-4960-449X}}
\author[2]{F. Nguyen\,\orcidlink{0000-0002-6332-7815}}
\author[4]{T. Niggemann}
\author[4]{N. Nippe}
\author[4,7]{J. Pretz\,\orcidlink{0000-0001-9128-419X}}
\author[4]{V. Tempel\,\orcidlink{0009-0004-3684-689X}}
\author[1]{B. Vink}
\author[1]{W. Walk}

\affil[1]{Nikhef, Science Park, 1098 XG Amsterdam, The Netherlands}
\affil[2]{University of Liège, Urban and Environmental Engineering, Liège 1 4000, Belgium}
\affil[3]{KU Leuven Kasteelpark Arenberg 40, Leuven, B-3001, Belgium}
\affil[4]{RWTH Aachen University, III. Physikalisches Institut B, Aachen, Germany}
\affil[5]{Maastricht University, 6211 LK Maastricht, The Netherlands}
\affil[6]{Royal Netherlands Meteorological Institute (KNMI), De Bilt, Netherlands}
\affil[7]{Institut f\"ur Kernphysik, Forschungszentrum J\"ulich, 52425 J\"ulich, Germany}

\date{}  

\begin{document}

\maketitle

\vspace{-1cm}

\begin{abstract}
The low-frequency sensitivity of the Einstein Telescope depends on the seismic environment of its site, both through the mechanical coupling of ground motion to the test masses and through Newtonian noise.
We present the first results of the seismic characterization of the Euregio Meuse-Rhine candidate site based on a network of recently instrumented boreholes covering the first seven months of 2026 at four underground stations and their co-located surface sensors. Two of these boreholes were deliberately sited next to railway infrastructure, so that the levels reported here describe how anthropogenic transients propagate to depth at different distances from the source. The analysis was carried out with \texttt{SiteChar}, a Python framework developed for this work to provide a common processing pipeline across different sensors, networks and observation periods. It processes the data in sequential chunks and separates the noise floor from the contribution of transients.
We present a characterization of the catalogue of observed transients and a tentative classification. 
These measurements form part of an integrated site characterization approach that combines the observed seismic field with three-dimensional geological and wave-propagation models to predict Newtonian noise and ultimately guide its mitigation at the Euregio Meuse-Rhine candidate site.
\end{abstract}

\section{Introduction}
\label{sec:intro}
The Einstein Telescope (ET) is designed to extend the observation band of gravitational wave detectors down to a few Hertz \cite{Punturo_2010,ETsensitivitystudies}.
This extension makes the characterization of the underground seismic environment a central element of site assessment.  The dedicated cryogenic low-frequency interferometer is designed to target approximately the 3--30~Hz band, enabling science including early warning for binary neutron-star mergers, observations of intermediate-mass black holes, and improved tests of gravity in the strong-field regime \cite{ETScience}. This is also the frequency band in which detector performance is most closely coupled to the local environment.

Below roughly 10~Hz \cite{SiteSelectionCriteria2020}, the seismic environment enters the sensitivity budget through two distinct mechanisms. First, ground motion couples mechanically to the test masses through the suspension chain. Seismic isolation attenuates this transmission, while underground siting provides an additional reduction by taking advantage of the lower seismic motion at depth. Second, fluctuations in the local mass density associated with the same ground motion exert a gravitational force directly on the test masses, called Newtonian noise (NN), which no isolation system can shield \cite{Harms2019}. Careful site selection can reduce this contribution, while active subtraction using arrays of auxiliary sensors provides an additional mitigation route. Current ET sensitivity projections anticipate the need for NN mitigation in the low-frequency detector, motivating substantial effort toward the optimisation of the auxiliary sensor arrays \cite{Driggers2012, Badaracco2019, Badaracco2020, Badaracco2024, Schillings2025, Schillings:2026ncw}, toward next-generation and alternative sensing technologies \cite{Coughlin2018, Ophardt2025, Rading2025}, and toward neural-network subtraction methods \cite{vanBeveren_2023, Kelleter:2026pep}. A prototype cancellation system has also been deployed at Virgo \cite{Koley2024a, Koley2024b} for testing. The central question for site characterization is therefore not simply how quiet a site is, but whether its seismic environment can be sufficiently characterized and modelled to predict its impact on ET and to design an effective NN mitigation system.

For the same reason, the comparison between candidate sites (the Sos Enattos area in Sardinia, Italy; the Euregio Meuse-Rhine (EMR) at the border between the Netherlands, Belgium and Germany; and the Lusatia site in Saxony, Germany) cannot be reduced to a comparison of ambient noise amplitudes. The conversion from ground motion to gravity fluctuations depends on the composition of the seismic wavefield, on its spatial coherence, and on the local geology, so that sites with comparable displacement spectra may differ appreciably in their NN projections, and viceversa. The relevant quantities are thus the wave content and coherence length of the field, its stability in time, the occurrence and characteristics of anthropogenic transients, and the extent to which all of these can be measured by an auxiliary array and used for subtraction.

An explicitly integrated approach to site characterization is crucial for 
the EMR candidate site. Rather than treating geology, seismic observations, and NN modelling as separate activities, the EMR strategy is to combine the measured seismic field with a detailed three-dimensional geological model and physics-based simulations of seismic wave propagation. The resulting displacement field can then be translated into gravitational perturbations at the detector test masses, including the influence of the underground infrastructure, and used to develop and optimize mitigation strategies. A distinctive feature of the EMR site characterization program is therefore the development of this integrated, site-specific modelling chain, connecting geological structure, measured seismic fields, wave propagation, NN prediction, and mitigation. This approach is designed to move beyond the characterization of individual site properties towards a physics-based prediction of how the complete underground environment affects ET performance. 

Previous underground campaigns in the EMR area have shown that the seismic field is dominated by a nearly continuous stochastic background of body waves with random incidence angles \cite{Bader_2022,Koley_2022}. Under the assumptions adopted in those studies, the resulting NN estimate lies above the ET design sensitivity by up to about a factor of two below $\sim 8$~Hz, with the largest margin required in the 3--6~Hz band. The same analysis indicate that increasing the depth beyond $\sim 250$~m would bring limited additional benefit, since the body-wave background is only weakly depth-dependent. These preliminary NN estimates depend on assumptions concerning the composition and spatial properties of the seismic field. The new borehole network provides substantially longer and more spatially distributed observations with which the site-specific models can be refined.

A further aspect of the seismic environment concerns its non-stationary component. Anthropogenic activity in the surrounding area produces transient disturbances that are superimposed on the ambient background, and whose contribution is not captured by long-term averaged spectra. Their characterization, in terms of spectral signature, occurrence statistics and spatial extent, is therefore a natural complement to the description of the background field and provides part of the input needed for a realistic description of the environment in which the detector would operate.

In this work, we show the first results from the recently deployed borehole network in the Terziet area to characterize the seismic environment of the EMR candidate site. We separately quantify the seismic background and the non-stationary transient population. Because two of the boreholes considered here were intentionally located close to railway infrastructure, these measurements should not be interpreted as representative noise levels at prospective detector vertices but they provide controlled observations of anthropogenic source generation, propagation and attenuation through the local geology. The resulting measurements provide constraints for the three-dimensional wave-propagation and NN models being developed for the EMR site.

An important consideration for NN modeling is that an analysis based primarily on vertical seismic spectra, as presented in this paper, provides relevant constraints on the seismic environment but cannot, on its own, reliably determine the seismic NN contribution.
For a diffuse, isotropic field with a known ratio of S- to P-wave energy, vertical borehole motion can constrain the horizontal acceleration of the test masses. Data from the TERZ sensor were recently used in this way to infer coincident NN transients in the triangular ET configuration \cite{harms2026correlatedcoincidentnoiseeinstein}. These assumptions must however be tested against measurements, especially in the presence of spatially varying transients, and realistic numerical models must reproduce both vertical and horizontal spectra. The characterization of the horizontal components and a full NN estimate are left to a future work.

To ensure consistent analysis across stations, sensors and observation periods, we developed the Python-based \texttt{SiteChar} framework, used throughout this work. It provides a common processing pipeline and allows to separate the seismic background from transient contributions.

The paper is organized as follows. Section~\ref{sec:stations} describes the site, the instrumentation and the dataset. Section~\ref{sec:sitechar} introduces the \texttt{SiteChar} package and the spectral estimators it defines. Section~\ref{sec:background} presents the background noise characterization, while Section~\ref{sec:transients} focuses on the transient population and on the coincident detection.  Section~\ref{sec:conclusion} summarizes the results and outlines future work.

\section{The EMR candidate area and station layout}
\label{sec:stations}

\begin{figure}[t!]
  \centering
\includegraphics[trim=1mm 1mm 1mm 1mm, clip, width=0.5\textwidth]{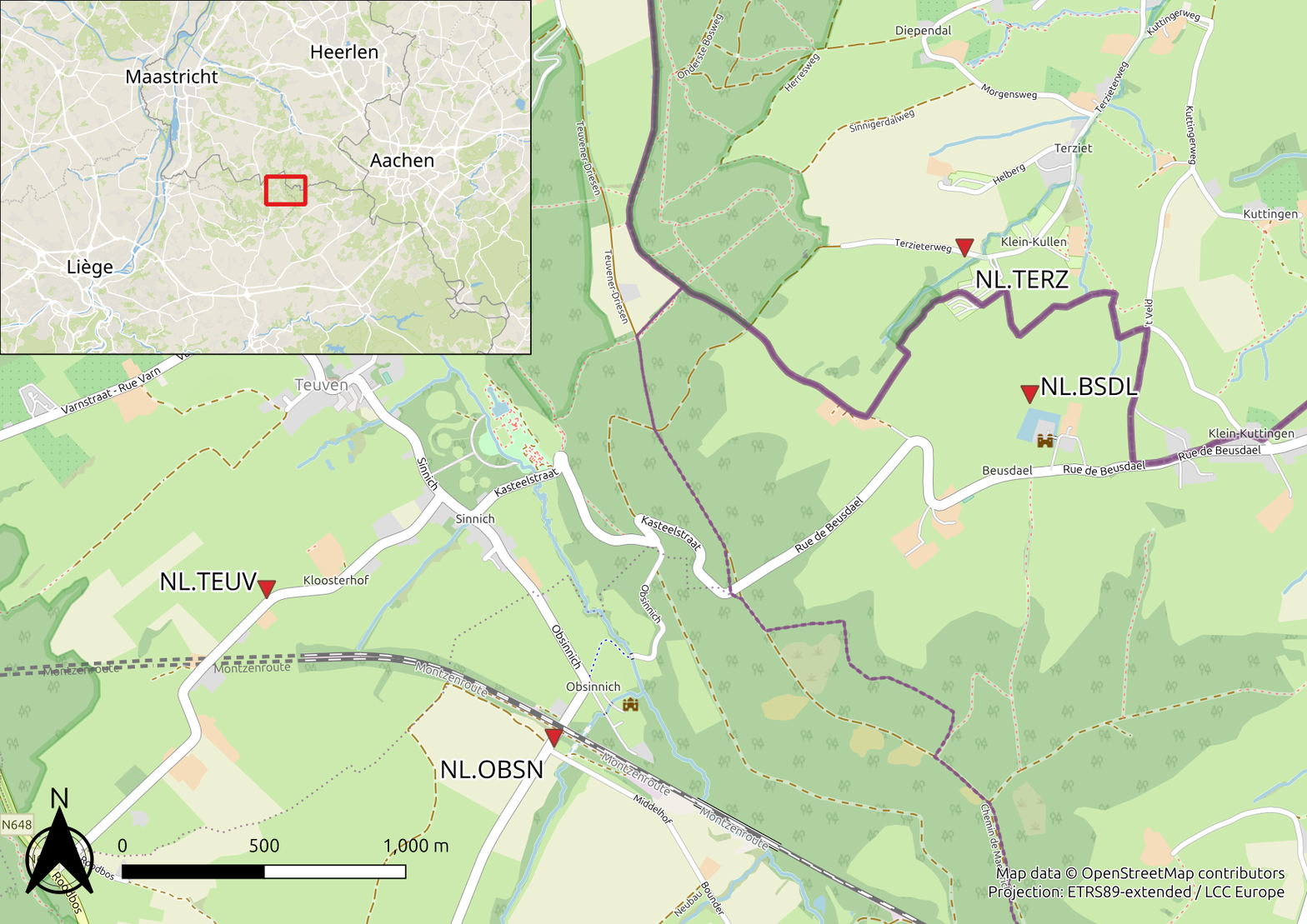}
  \caption{Map of the Terziet region showing the installed stations (red triangles) and the main transport infrastructure in their vicinity.}
  \label{fig:map}
\end{figure}

The EMR candidate area lies at the border between the Netherlands, Belgium and Germany. 

Long-term reference observations in the area are provided by the underground station at Terziet (TERZ), operating at 250~m depth. As part of the ongoing exploration effort, a set of recently drilled boreholes has been instrumented with broadband seismometers at the surface and borehole, at Obsinnich (OBSN), Teuven (TEUV), and Beusdael (BSDL) as shown in Figure \ref{fig:map}. Further boreholes at Hoof, Asse, Herbesthal are being completed and instrumented, and are not covered by the present analysis.

\begin{figure}[t!]
  \centering
\includegraphics[width=0.65\textwidth]{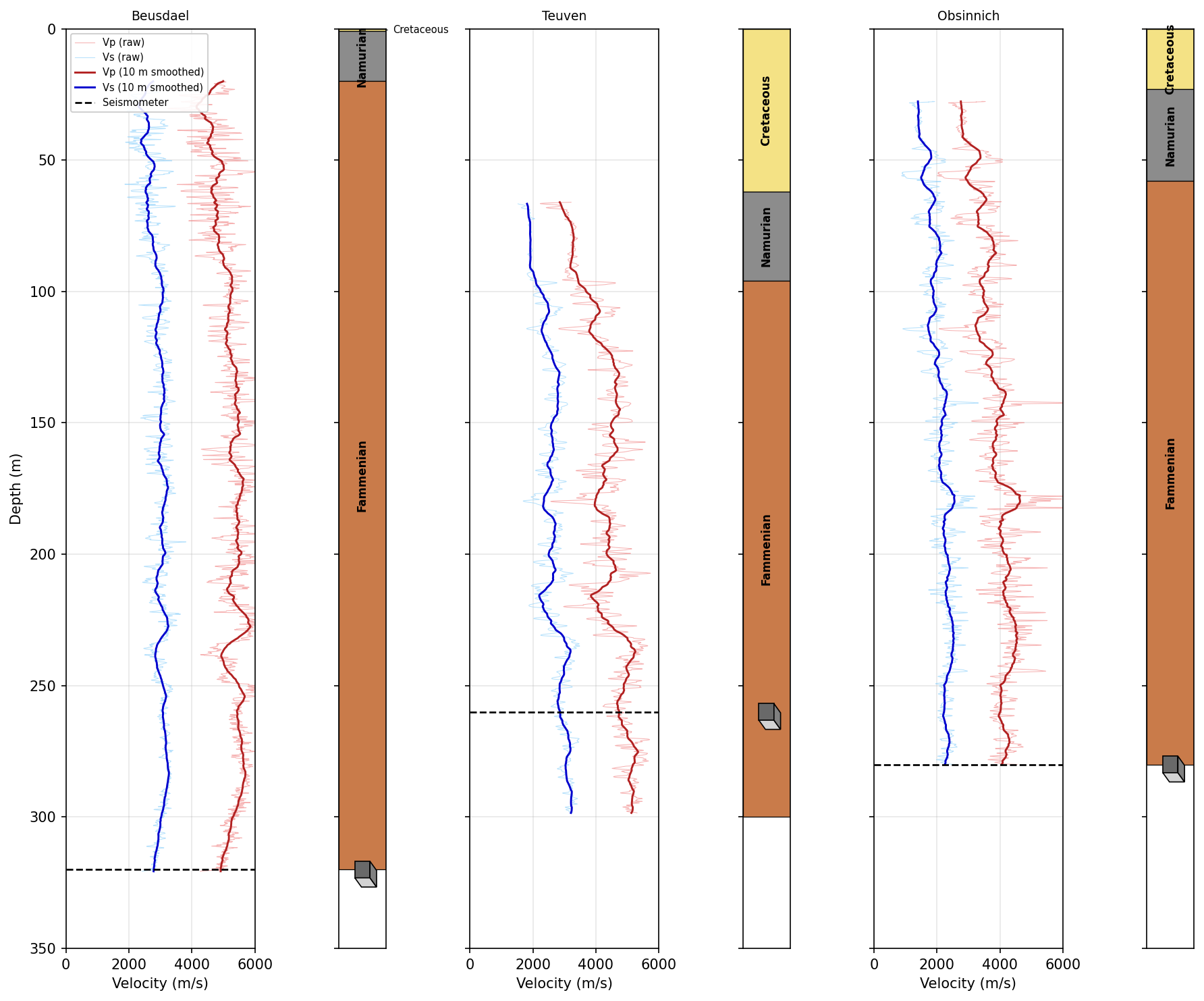}
  \caption{Geological logs of the boreholes in (from left to right) BSDL, TEUV and OBSN, together with their sonic logs.}
  \label{fig:logs}
\end{figure}

The boreholes in TEUV, OBSN and BSDL are all exploring the Boze-Le Val-Dieu ridge, a geological structure bringing upwards the Famennian rocks. In this area, the Dinantian limestones, normally found between the Namurian shales and the Fammenian sandstones, are absent \cite{poty1991tectonique}. This was observed in the Terziet borehole \cite{Koley_2022}, but is confirmed in the new boreholes. They all show a sequence with largely unconsolidated Cretaceous sediments in the first meters, followed by a decametric succession of shales from the Namurian, before the target geological unit, the Famennian sandstone (see Figure \ref{fig:logs}).  In the sonic logs from TEUV and OBSN, a sharp contrast in velocities between the Namurian shales and the Fammenian sandstones is observed.

It should be emphasised, and it is essential for the interpretation of all the results presented below, that two of these boreholes were \emph{deliberately} positioned in unfavourable locations (see Figure \ref{fig:map}). Of the stations considered here, only BSDL is close to an envisaged corner point of the detector. OBSN and TEUV were chosen specifically to study the generation, propagation and attenuation of anthropogenic noise in the local geology, and therefore do not represent the noise conditions expected at candidate corner-point positions. The relevant characteristics for each station are:

\begin{itemize}
  \item \textbf{Obsinnich (OBSN)} is located about 60~m from an active
    railway line and railway bridge, with train traffic throughout the day and
    night. The borehole sensor is installed at 280~m. Additional structures in
    the vicinity include the Viaduct of Remersdaal, roughly 530~m to the east,
    whose pillars are founded in the bedrock, and a tunnel portal about 940~m
    to the west.
  \item \textbf{Teuven (TEUV)} lies approximately 200~m from a
    railway tunnel, again with traffic distributed over the whole day, with the
    borehole sensor at 260~m.
  \item \textbf{Beusdael (BSDL)} is the only station of this set close to a
    possible corner-point location. It is instrumented at three levels: at the surface, at 30~m and at 320~m. The deepest sensor was only properly seated at its target depth on 26 May 2026, after a series of installation problems, and since then only the vertical and one horizontal component are operational. In this work we therefore consider exclusively the
    320~m sensor for the time window following its final installation. 
\end{itemize}

The surface sensors are Güralp 3ESPC three-component broadband seismometers, while the borehole sensors, at both intermediate and total depth, are Nanometrics Trillium Compact Post-Hole 120~s three-component broadband seismometers. At Terziet, a malfunctioning Kinemetrics STS-5 was replaced by a Trillium Compact Post-Hole unit, which has been recording since 11 November 2025; the power spectral densities obtained with the two configurations are in agreement, so that the continuity of the reference dataset is preserved. Data are recorded continuously at a sampling rate of 40~Hz; all spectra shown in this work are corrected for the instrument response, expressed in acceleration units and quoted as $10\log_{10}\left[S/S_0\right]$ with $S_0 = 1\,\mathrm{(m/s^2)^2/Hz}$ where $S$ is the power spectral density (PSD) of the seismic field; for example, a PSD of $-140$~dB corresponds to $S = 10^{-14}\,\mathrm{(m/s^2)^2/Hz}$. 

The analysis strategy adopted throughout this work and the definitions of the spectral quantities used are described in the next section.

\section{The \texttt{SiteChar} package: a general-purpose framework for site characterization}
\label{sec:sitechar}

Characterizing the seismic environment of a candidate site is a necessary step in site selection and monitoring towards the ET site selection. In practice, this work is often repeated independently for each station and candidate site, using ad-hoc scripts for the seismic characterization. This duplication of effort makes results difficult to compare across stations and sites and slows down the iterative process of site evaluation. \texttt{SiteChar} was developed to address this: a single, general-purpose analysis framework that separates the \emph{scientific question} being asked (how noisy is this site, how often do transients occur, how do they propagate between sensors) from the \emph{specifics of any one deployment} (sensor type, data format, network convention, deployment duration).

\texttt{SiteChar} is deliberately agnostic to the physical nature of the sensor and to the geographic scale of the deployment. This generality is achieved by defining the framework around a minimal set of physical concepts: a sensor, a station grouping co-located sensors, and a network grouping distinct stations for comparison. The same analysis can then be applied uniformly to a single borehole seismometer, to a pair of co-located surface and borehole sensors at one station, or to a set of geographically distant stations belonging to entirely different monitoring networks. A network can equally group stations belonging to different candidate sites, which are then processed together, in a single run and with identical settings. Adding a new site, or a new class of sensor beyond seismometers (e.g. magnetometers, microbarometers), requires only implementing the data-loading step for that source, after which the analysis routines are used unchanged.

For long-term characterization, \texttt{SiteChar} processes the data in sequential chunks (typically two hours), retaining only the derived spectral products needed for characterization and discarding the raw samples once each chunk has been processed. This makes multi-month characterization runs practical on standard hardware.

Transients (e.g.\ passing trains, vehicle traffic, or other anthropogenic disturbances) are identified from the time-frequency structure of the signal. For each analysis chunk, a spectrogram is computed and, at each frequency, the power is converted to a robust z-score relative to the median and median absolute deviation across time (see details in Section \ref{sec:zscore}). A time sample is flagged as part of a transient if its z-score over a chosen frequency band of interest exceeds a specific threshold for a minimum duration; adjacent flagged samples are merged into a single detected event. 

For every analysis chunk, \texttt{SiteChar} defines three different spectral quantities:

\begin{itemize}
  \item The \textbf{chunk spectrum} is the power spectral density computed
    over the entire analysis chunk (typically two hours), with no
    exclusion of transients. It answers the question ``what does the
    ambient seismic field look like overall, without post-processing, averaged
    over this period?''.

  \item The \textbf{background spectrum} is computed only over the longest
    transient-free interval identified within the same chunk, using the
    transient detection to exclude all flagged intervals (with a guard margin around each). It
    answers a different question: ``what is the seismic field like in the
    absence of the fast disturbances that were detected?''.

  \item The \textbf{event spectrum} is computed over an individual detected
    transient, and is always paired with the background spectrum of the
    same chunk to form an event-to-background ratio. This ratio
    quantifies how much a given transient elevates the noise level above
    the ambient background, as a function of frequency.
\end{itemize}

The chunk and background spectra are not intended to replace the spectral estimates adopted so far in ET site characterisation, such as spectral histograms and PPSDs computed over all available data, which remain the reference for comparing sites and for quantifying the time spent at a given noise level. They must be considered diagnostic quantities tied to the transient detection: the background spectrum provides the reference against which individual events are measured, and the difference between chunk and background spectra quantifies the contribution of the detected transients to the average noise level.

Alongside the spectra, each processed chunk contributes its detected transients to a catalogue, in which every entry carries the times of the trigger and of the peak of the detection statistic, the equivalent duration defined in Section~\ref{sec:zscore}, the in-band energy, and the event spectrum paired with the background of its own chunk. The catalogue is what makes two different levels of analysis possible with the same processing. A single entry can be expanded back into the full analysis of one transient, reloading the chunk it comes from and adding sensors that were not part of the original run, such as the co-located surface instruments used in Section~\ref{sec:single_event}. The catalogue as a whole can instead be treated as a population, whose distributions in duration, energy, hour of the day and coincidence between stations are the subject of Sections~\ref{sec:catalogue} and~\ref{sec:coincidence}.

\subsection{Z-score}
\label{sec:zscore}

\begin{figure}[t!]
\centering
\includegraphics[width=0.8\textwidth]{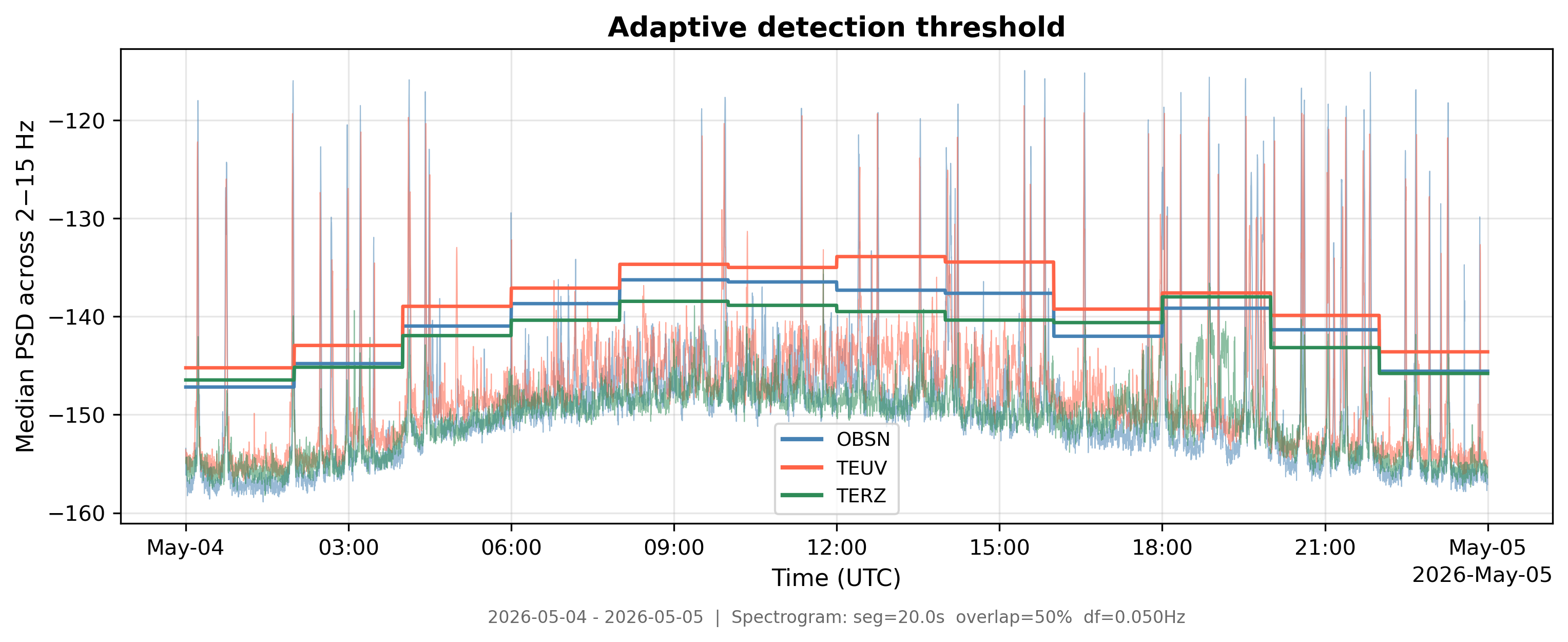}
\caption{Detection threshold in physical units for the analysis of one day (4 May 2026). Thin lines: median power spectral density across the 2–15 Hz band, one value per 10 s spectrogram column, for the three underground vertical channels. Thick lines: the corresponding detection threshold, obtained by re-estimating the per-frequency median and median absolute deviation on each 2 h block and applying the fixed z-score threshold z = 2.5 used throughout this work. The vertical distance between the two lines of a given colour is the detection statistic itself, at $\sim 3.5$ dB per unit z. Because the normalisation is re-estimated locally, a fixed z corresponds to an absolute level that follows the diurnal cycle of the ambient noise, spanning about $5--8$ dB between night and midday.
}
\label{fig:adaptative_threshold}
\end{figure}

Transients are identified independently on each sensor from a normalized spectrogram statistic. The record is processed in consecutive blocks of two hours, each loaded with $600$~s of padding on either side so that the transients of the response deconvolution fall outside the analyzed interval, and transformed with a Hann-tapered short-time Fourier transform using $20$~s segments and $50\%$ overlap, giving $\Delta t = 10$~s and $\Delta f = 0.05$~Hz. Writing $S(f,t)$ for the resulting power spectral density, each frequency row is standardised by its own median $m(f) = \operatorname{med}_t \log_{10} S(f,t)$ and median absolute deviation $\mathrm{MAD}(f) = \operatorname{med}_t |\log_{10} S(f,t) - m(f)|$, both evaluated over all columns of the block.
\\The detection statistic is the median of the standardized spectrogram across the analysis band $[f_1,f_2]=[2,15]$~Hz:
\begin{equation}
  z(t) = \operatorname*{med}_{f_1 \le f \le f_2} \frac{\log_{10} S(f,t) - m(f)}{\mathrm{MAD}(f)} .
  \label{eq:score}
\end{equation}

Columns with $z(t)$ above the threshold $z_{\rm thr}$ are flagged and contiguous flagged columns are merged into triggers. Since $\log_{10} S$ is a logarithmic power, the statistic converts to physical units through a single factor: an excursion $z$ corresponds to an excess of $10\,\overline{\mathrm{MAD}}\,z$ dB over the local background, where $\overline{\mathrm{MAD}} = \mathrm{med}_{f_1 \le f \le f_2} \mathrm{MAD}(f)$.  

The construction assumes that the ambient noise is stationary within a block, that transients occupy a minority of it so that the median and the median absolute deviation still describe the stationary state, and that the signals of interest are broadband, since the median in \eqref{eq:score} responds only when more than half of the frequency bins rise together. The median absolute deviation is used unscaled, so $z_{\rm thr}$ is an empirical operating point and not a Gaussian tail probability. 

Because the normalisation is re-estimated on every block, a fixed threshold in $z$ does not correspond to a fixed physical amplitude: the absolute level required to trigger follows the diurnal cycle of the ambient noise, varying by approximately $10$~dB between night and midday (Figure \ref{fig:adaptative_threshold}), so the catalogue is not amplitude-complete in an absolute sense and weaker transients are recovered at night but missed during the day. 

In this work, we use a threshold $z_{\rm thr} = 2.5$. For the analyzed data in this campaign, $10\,\overline{\mathrm{MAD}} \simeq 3.5$ dB; it follows that $z_{\rm thr} = 2.5$ corresponds to an excess of about $8$--$9$ dB with respect to the noise PSD floor, as shown in Figure \ref{fig:adaptative_threshold}.

\begin{figure}[t!]
  \centering
\includegraphics[width=0.6\textwidth]{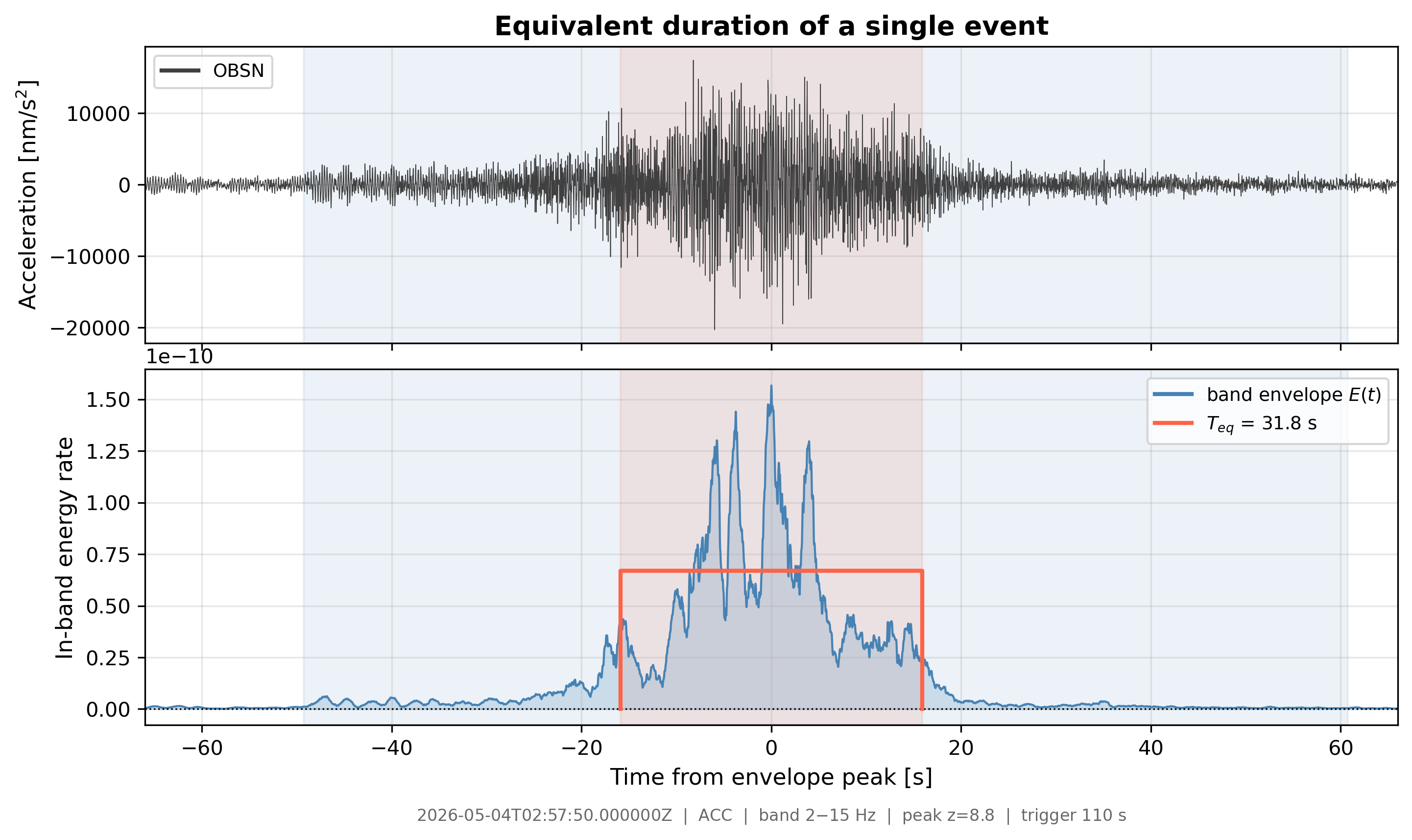}
  \caption{Equivalent duration of a single event at station OBSN. Top: ground acceleration. Bottom: in-band energy rate $E(t)$, the squared analytic envelope of the $2$--$15$~Hz signal smoothed over $1$~s, with the boxcar of equal area, of width $T_{\mathrm{eq}}=(\int E)^2/\int E^2$ and height $\int E^2/\int E$, centred on the envelope peak. The blue band marks the trigger returned by the detector, the red band the $T_{\mathrm{eq}}$ integration window.}
  \label{fig:teq}
\end{figure}

\subsubsection{Duration of a trigger}

A trigger is a contiguous set of spectrogram columns with $z(t)>z_{\rm thr}$, but its extent depends on the signal amplitude and therefore does not directly measure the transient duration. 

We therefore use the equivalent duration of the band envelope. The instrument-corrected acceleration is band-passed to $2$--$15\,$Hz with a fourth-order Butterworth filter. The  instantaneous power envelope $E(t)$ is obtained from the Hilbert transform, smoothed over $1\,$s, and corrected for the local baseline $E_0$. Defining
\begin{equation}
    \tilde{E}(t)=\max\left[E(t)-E_0,\,0\right],
\end{equation}
the equivalent duration is
\begin{equation}
    T_{\rm eq} =
    \frac{\left[\displaystyle\int \tilde{E}(t)\,\mathrm{d}t\right]^2}
         {\displaystyle\int \tilde{E}^2(t)\,\mathrm{d}t},
    \label{eq:teq}
\end{equation}
evaluated over a window of half-width
$\max(3\,\tau_{\rm trig},150\,{\rm s})$ centred on the trigger where $\tau_{\rm trig}$ is the trigger duration. This measure is independent of amplitude for a fixed envelope shape and is not limited by the spectrogram cadence. The equivalent duration $T_{\rm eq}$ can be defined as the duration of a rectangular pulse with the same energy and effective power as the observed transient (see Figure \ref{fig:teq}).

Because the search window has a minimum half-width of $150\,$s, neighbouring triggers may contaminate the event window. We therefore record the distance to the nearest trigger at the same station and flag events as isolated when no neighbour falls inside the window.

\section{Background seismic noise characterization}\label{sec:background}

The analysis covers the period from 1 January 2026 to 1 August 2026, processed in sequential two-hour chunks as described in Section~\ref{sec:sitechar}. For each chunk, both the chunk spectrum and the background spectrum are computed, so that the stationary noise floor can be separated from the contribution of the detected transients.

Throughout this work only the vertical component is analyzed. The azimuthal orientation of the borehole horizontal components is not independently constrained, so the horizontal traces cannot yet be rotated into a common reference frame for a station-to-station comparison. Orientations can in principle be recovered from the polarisation of earthquakes, and this is left to a forthcoming study. The horizontal components carry information that is essential for NN estimation and cannot be neglected in the long term.

\begin{figure}[t!]
\centering
\includegraphics[width=\textwidth]{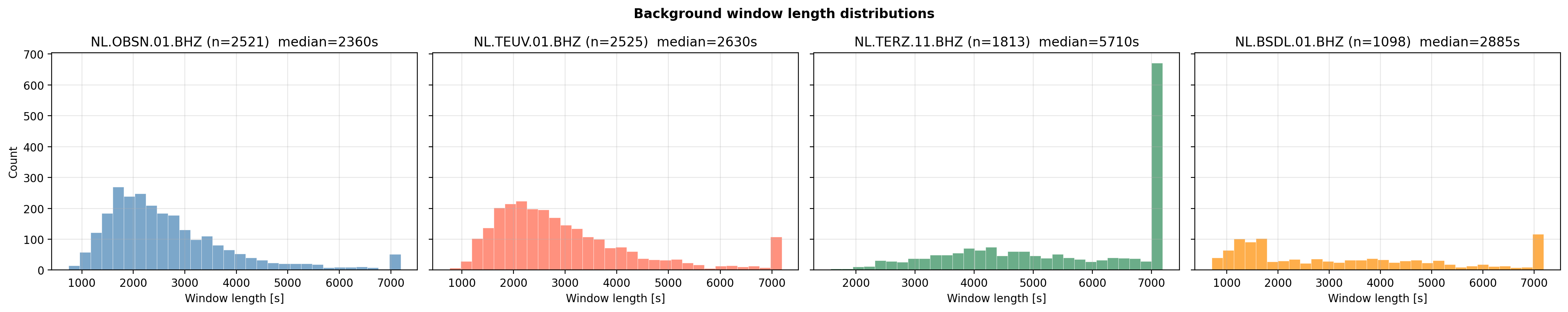}
\caption{Distribution of the background window length at each station, i.e.\ the duration of the longest transient-free interval found in each two-hour chunk. The lower edge is set by the 400~s minimum required for the spectral estimate; the excess at 7200~s corresponds to chunks in which no transient was detected. In practice, the minimum length for OBSN, TEUV and BSDL is 900~s while for TERZ is 1900~s.}
\label{fig:bg_window_lengths}
\end{figure}

The background spectra discussed in this section are computed on the longest transient-free interval of each two-hour chunk, as defined in Section~\ref{sec:sitechar}. The duration of that interval is not a fixed parameter but a property of the station, set by its local transient rate: it ranges from the 400~s minimum required for the Welch estimate with the chosen parameter ($100$~s length with $50\%$ overlap) up to the full 7200~s of a chunk with no detections. Figure~\ref{fig:bg_window_lengths} shows that the medians of the background window duration differ by more than a factor of two across the network, from 2360~s at OBSN to 5710~s at TERZ, with the pile-up at 7200~s collecting the transient-free chunks; the four background spectra are therefore averaged neither over the same amount of time nor over coincident time slots within each chunk.

\begin{figure}[t!]
\centering
\includegraphics[width=\textwidth]{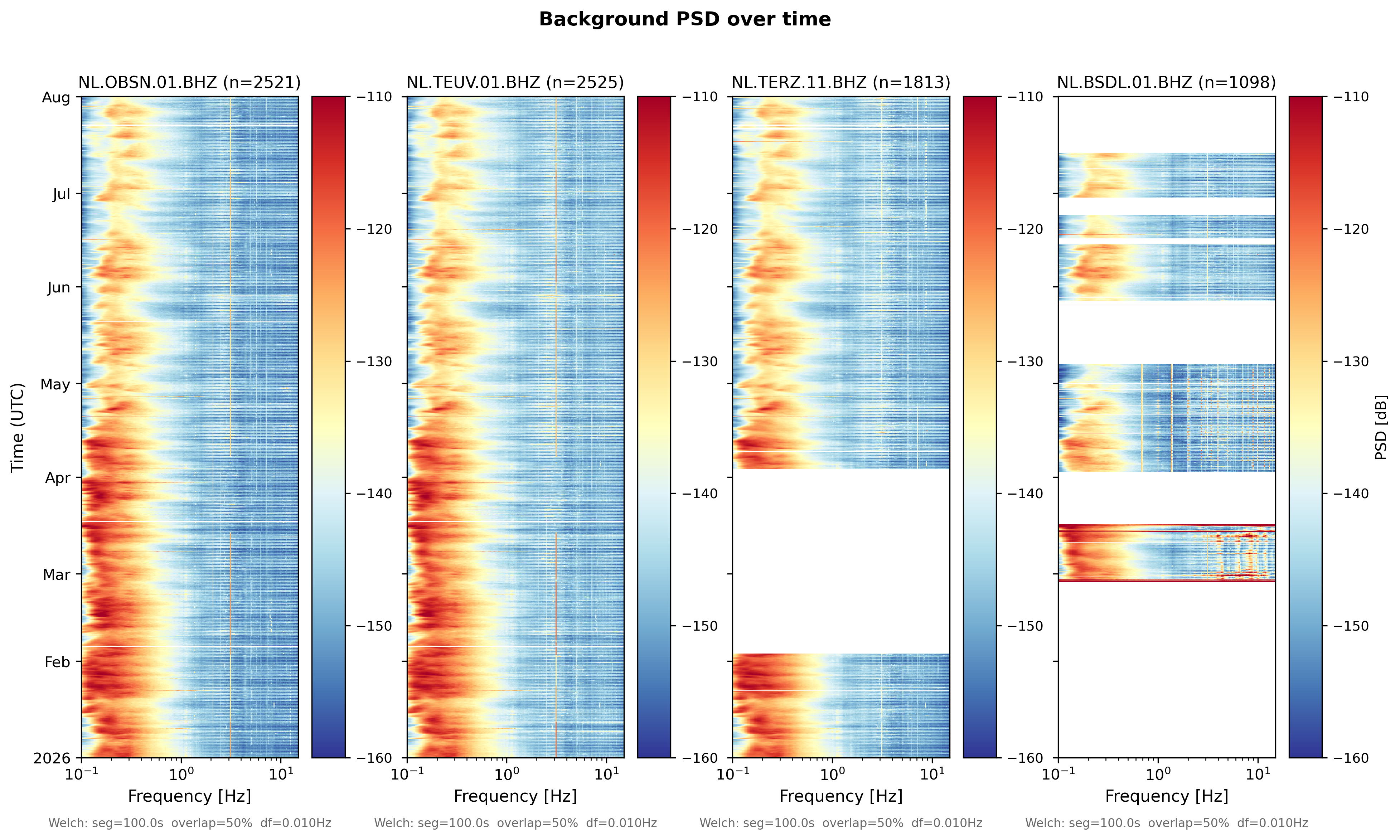}
\caption{Time evolution of the background PSD at each borehole station over the January--August 2026 analysis window. Each column corresponds to one station; the vertical axis is calendar time (UTC), the horizontal axis is frequency on a logarithmic scale. Colour encodes PSD level in dB. White regions indicate gaps in data coverage. Seasonal and diurnal variations in the noise floor are visible in all the stations.}
\label{fig:heatmap_bg_psd}
\end{figure}

The number of chunks effectively available differs from station to station. The missing coverage is not uniformly distributed in time (for example, TERZ has a two-month gap between 3 February and 3 April) so that the four datasets sample different fractions of the seasonal cycle for the microseism activity for example. The BSDL borehole sensor requires a separate remark. The channel is available from 11 February, but the first two months are dominated by the residual disturbances of the installation, visible in Figure~\ref{fig:heatmap_bg_psd} as an unstable and elevated noise floor with spectral lines, and are therefore not representative of the site. Only the chunks after 26 May 2026 are used in the following.

Figure~\ref{fig:heatmap_bg_psd} shows the time evolution of the background PSD over the analysis period, with a resolution of two hours since each row is the background spectrum of one chunk. The microseismic band exhibits the expected seasonal decay, with levels progressively decreasing from January--March towards May--June by $\sim 10$~dB, in agreement with the North Atlantic wave forcing (Appendix~\ref{app:microseism}). A diurnal modulation of the noise floor is visible at all stations above 1~Hz with a PSD modulation of 5--8~dB in the 2--15~Hz band between the nighttime minimum and the midday maximum. This modulation of the background, and not of the sources, is what makes the detection statistic of Section~\ref{sec:zscore} intrinsically relative and is used in Section~\ref{sec:catalogue} to interpret the hourly distribution of the detections.

\begin{figure*}[t!]
\centering
\begin{subfigure}{0.7\textwidth}
    \includegraphics[width=\linewidth]{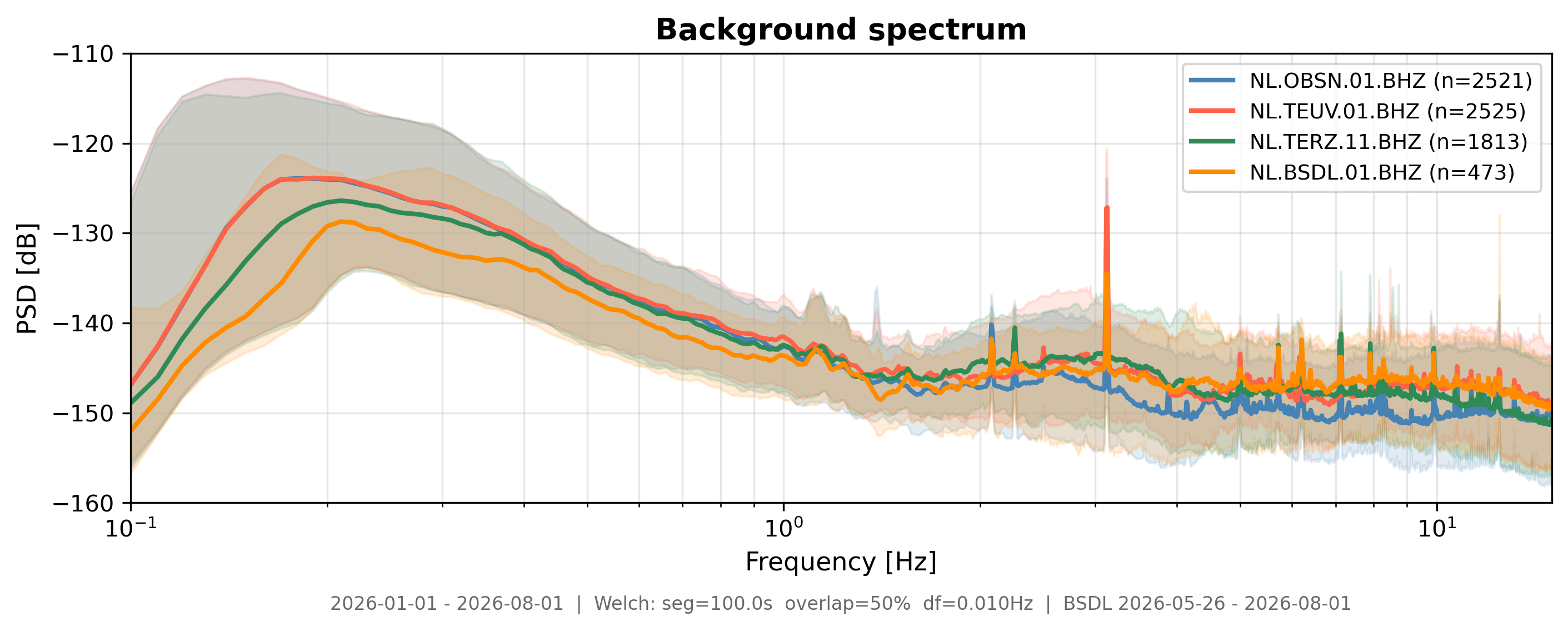}
\end{subfigure}
\hfill
\begin{subfigure}{0.7\textwidth}
    \includegraphics[width=\linewidth]{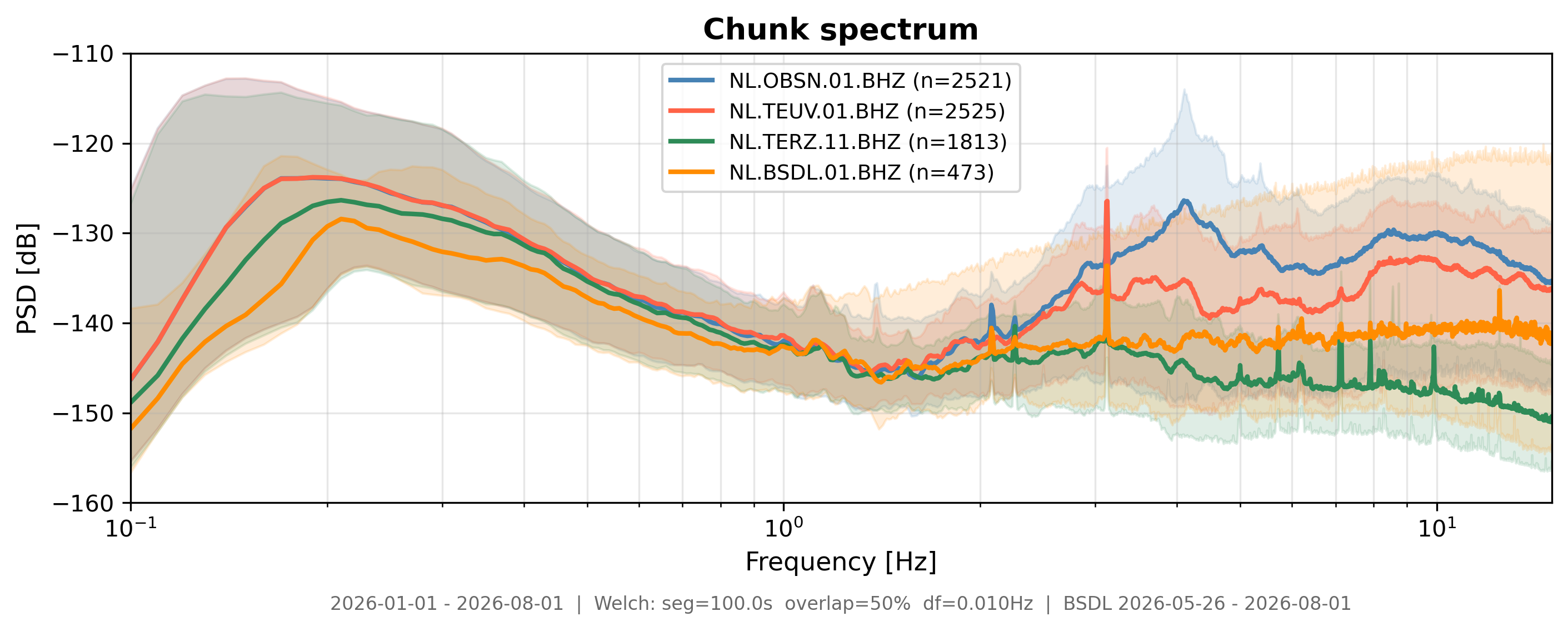}
\end{subfigure}
\caption{PSD comparison across the four underground borehole stations (OBSN, TEUV, TERZ, BSDL) over the January--August 2026 analysis window. Solid lines show the median, shaded bands the 5\textsuperscript{th}--95\textsuperscript{th} percentile range. \textit{Top:} background noise estimated from the longest quiet stretch in each 2-hour chunk. The background spectra use 37\% (OBSN), 42\% (TEUV), 77\% (TERZ) and 65\% (BSDL) of the available data.  \textit{Bottom:} full-chunk PSD including transients. The difference between the two panels reflects the spectral contamination introduced by transient sources, predominantly above 1\,Hz. For BSDL, we considered only the data after the 26 May.}
\label{fig:psd_overlay}
\end{figure*}

Two families of features stand out from the diffuse background. Horizontal stripes, short-lived and broadband in the microseism band, correspond to teleseismic events; their identification and their signature in the detection statistic are discussed in Appendix~\ref{app:earthquakes}. Vertical stripes, narrow in frequency and persistent over the whole observation period, correspond to spectral lines: the most prominent one is the line at $\sim 3.125$~Hz, clearly visible at OBSN and TEUV and weaker at TERZ and BSDL, which also appears as a sharp spike in the median spectra of Figure~\ref{fig:psd_overlay}. The amplitude of these lines is not stationary and their origin has not yet been established; candidate explanations include nearby rotating machinery, pumping or ventilation equipment.

Figure~\ref{fig:psd_overlay} (top panel) shows the median background PSD and the $5^{\rm th}$--$95^{\rm th}$ percentile range for the four borehole sensors. Below $\sim 0.5$~Hz the spectra of all stations are dominated by the secondary microseismic peak at $\sim 0.2$~Hz, with median levels between $-124$ and $-117$~dB. The apparent spread of a few dB between stations is not a property of the sites but a consequence of the different temporal coverage discussed above, since the microseismic amplitude is modulated by the seasonal cycle of ocean-wave activity. When the comparison is restricted to the time slots common to all stations the peaks coincide to within a fraction of a dB (see Figure~\ref{fig:common_chunks} in appendix). Between $\sim 0.5$ and $\sim 1$~Hz all sensors converge towards a common minimum of about $-148$~dB, and above 1~Hz they remain within a few dB of one another, OBSN defining the lowest floor at about $-150$~dB.

The main result of this comparison is therefore the similarity of the four curves that it is used as a sanity check for the detection procedure. Once the fast non-stationarities are excluded, four boreholes drilled in similar geological conditions, at depths between 250 and 320~m, and at distances from the railway ranging from tens of metres up to about two kilometres, describe essentially the same background field. It should be stressed that the exclusion concerns only the fast non-stationarities: the slow ones survive in these spectra, since each curve is a median over chunks that span the diurnal cycle, the weekly cycle and the seasonal modulation of the microseism, all of which contribute to the width of the percentile bands.

The chunk spectra of the bottom panel behave in the opposite way: the stations separate clearly above 1~Hz, with the two sited next to the infrastructure, OBSN and TEUV, lying well above the other two. At TERZ the two estimators differ by 1 dB at most, so its chunk spectrum is essentially its background; at BSDL the difference is instead driven by the short-duration population discussed in Section~\ref{sec:shorts}. The separation seen in this panel is therefore produced by the transient population and by its duty cycle, and not by the sites themselves. Reported together, the two panels separate the ambient level of a station from the frequency and amplitude of its perturbations.

\section{Transient event characterization at EMR}
\label{sec:transients}

In this section we describe the characterization of the transient catalogue detected in the first seven months of 2026 using the $z$-score statistic defined in Section~\ref{sec:zscore}, evaluated in the 2--15~Hz band with a threshold $z_{\rm thr}=2.5$ and a minimum duration of 10~s.  Section~\ref{sec:single_event} describe one coincident transient in detail using the four underground sensors and the co-located surface sensors. Section~\ref{sec:catalogue} describes the catalogue as a whole while Section~\ref{sec:coincidence} uses the coincidence between stations to separate the populations that compose it.

\subsection{Analysis of a single transient}
\label{sec:single_event}

\begin{figure}[t!]
\centering
\includegraphics[width=0.7\textwidth]{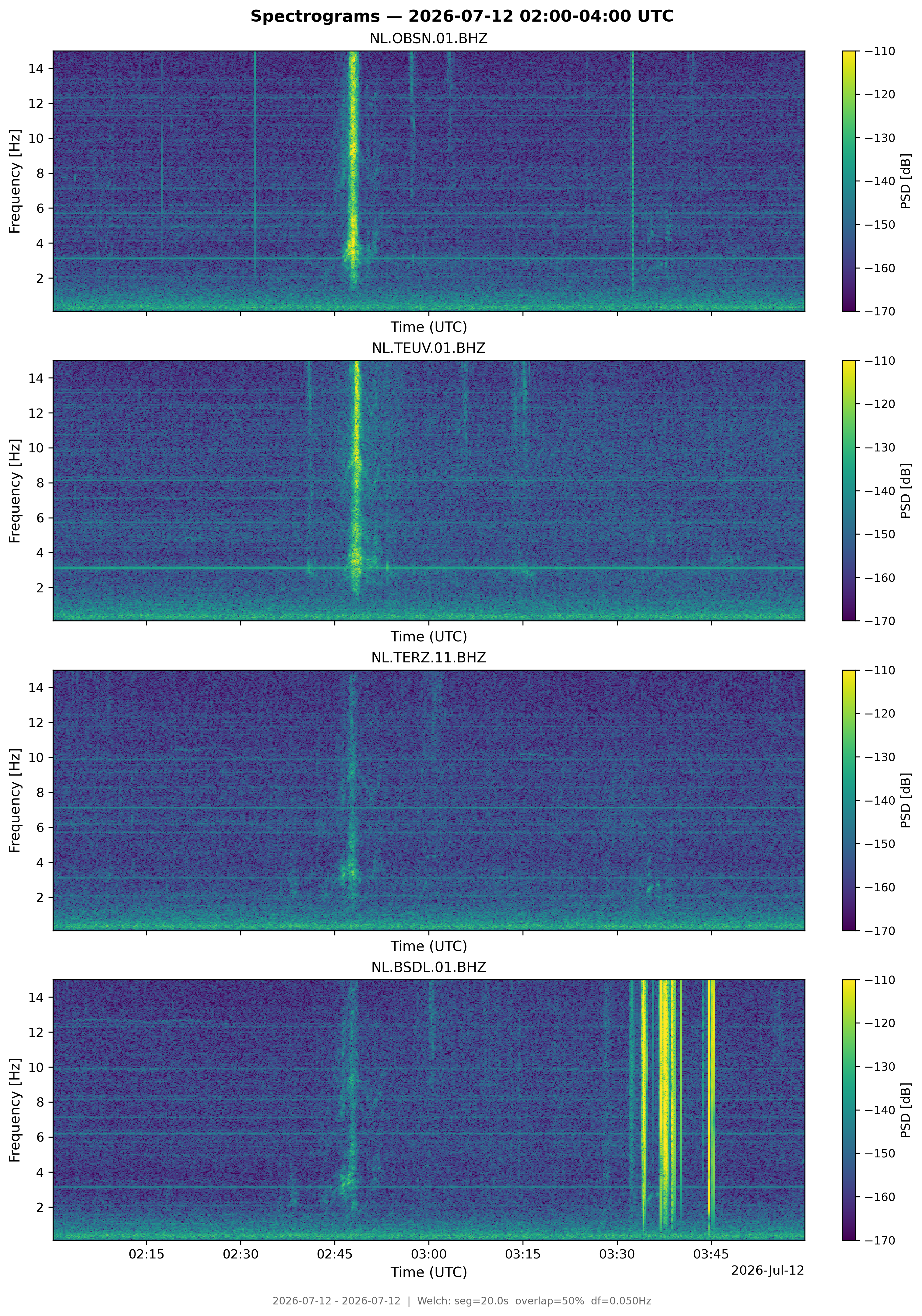}
\caption{Stacked spectrograms of the two-hour analysis chunk (02:00--04:00 UTC, 2026-07-12) at the four borehole sensors. The passage analyzed in Section~\ref{sec:single_event} appears as the broadband vertical burst near 02:48~UTC, visible at all four stations with station-dependent amplitude. The continuous horizontal line at $\sim 3.1$~Hz is the spectral line discussed in Section~\ref{sec:background}, unrelated to the passage.}
\label{fig:spectrograms_stacked}
\end{figure}

The single-event analysis is presented for a passage detected at the four borehole stations on 2026-07-12 around 02:48~UTC. The event is one entry of the catalogue, and the window analyzed is the two-hour chunk it comes from. For the single-event analysis we consider also the surface sensor of each station.

\begin{figure}[t]
  \centering
  \begin{subfigure}[t]{0.49\textwidth}
    \centering
    \includegraphics[width=\linewidth]{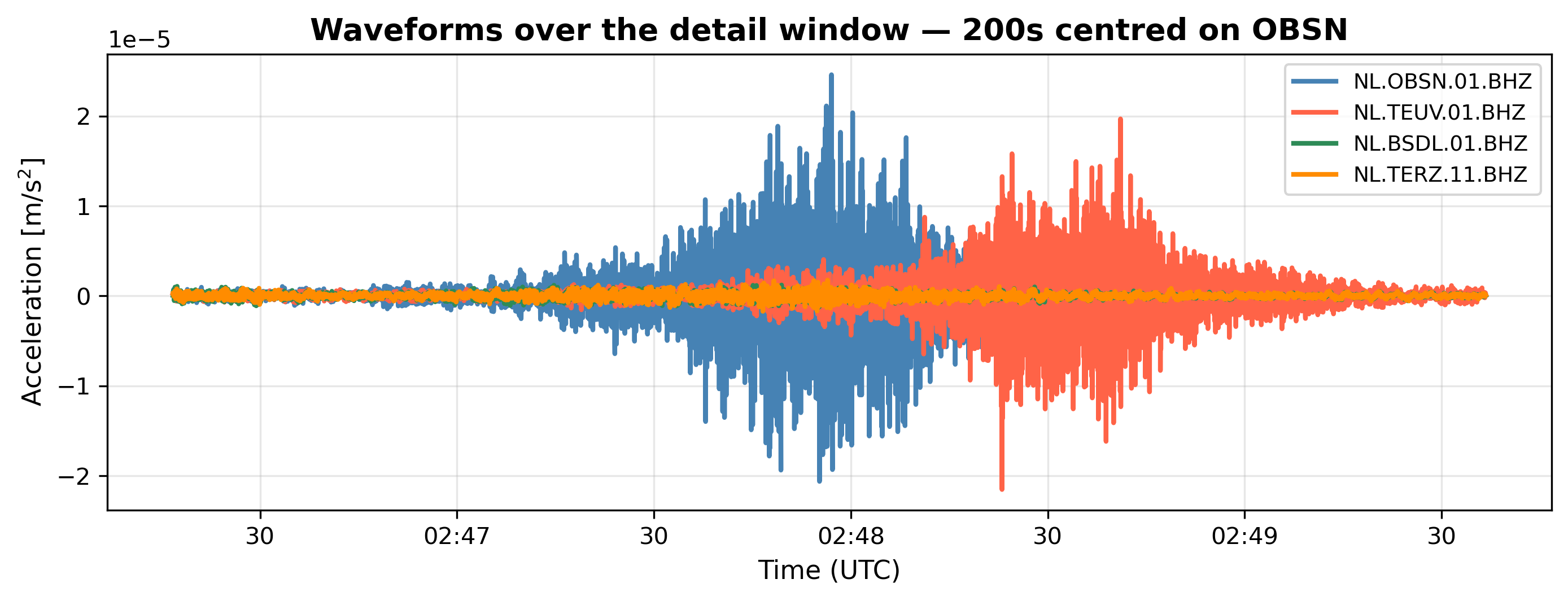}
    \caption{}
    \label{fig:waveform}
  \end{subfigure}
  \hfill
  \begin{subfigure}[t]{0.49\textwidth}
    \centering
    \includegraphics[width=\linewidth]{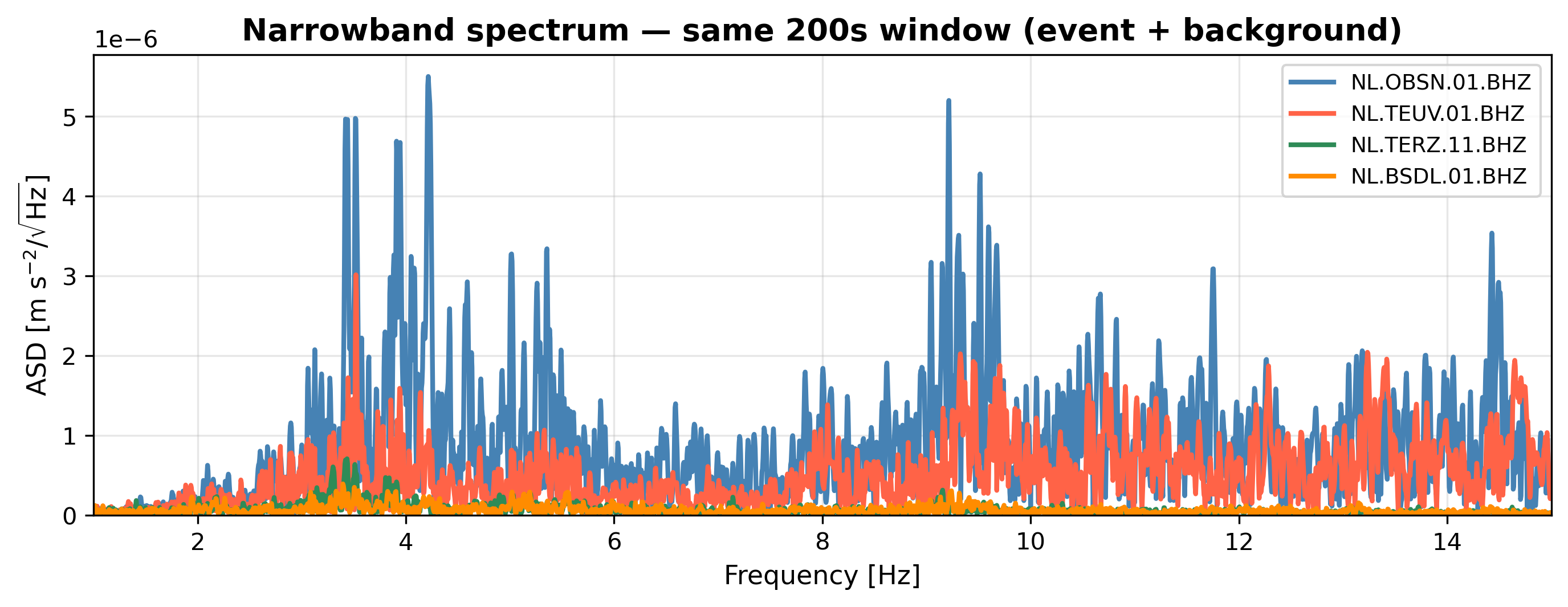}
    \caption{}
    \label{fig:narrowband}
  \end{subfigure}
  \caption{The passage shown in Figure \ref{fig:spectrograms_stacked} seen on a single 200\,s window common to the four stations, centred on the envelope peak at OBSN. \textit{Left:} instrument-corrected acceleration, drawn in order of decreasing $z$. \textit{Right:} amplitude spectral density on the same window, obtained from a single Hann-tapered FFT with no averaging, so that $\Delta f = 1/T$ with $T=200$s. This panel is intended to resolve spectral lines and not to read noise levels.}
  \label{fig:waveform_narrowband}
\end{figure}

Two windows are used throughout. The \emph{event window} is $T_{\rm eq}$--wide and centred on the maximum of the 2--15~Hz energy envelope, which is the same window on which the detection measures $T_{\rm eq}$ defined per station. The \emph{background window} is the longest contiguous stretch of the chunk in which \emph{neither} the surface and the underground sensor of the station has a trigger, with a guard of 60~s around each. The requirement on both sensors is essential: the surface detects considerably more events than the borehole, so a stretch that is quiet for the borehole is not necessarily quiet at the surface, and a contaminated reference inflates the measured attenuation. 

The surface sensor is counted as having recorded the event if its $z$-score exceeds the detection threshold inside the $T_{\rm eq}$ window of the borehole. Where the surface does not cross the threshold the event is not measurable, but it is bounded from above, since its surface PSD is at most the surface background. This case is not rare and is itself a result: at TERZ and BSDL the passage remains above the borehole noise floor while falling below the surface one, which is 20--30~dB higher, so that the gain in detectability provided by the depth can be quantified event by event.

Figure~\ref{fig:spectrograms_stacked} shows the two-hour data at the four borehole sensors. The transient appears as a broadband vertical burst, almost simultaneous within the resolution of the spectrogram and with station-dependent amplitude. The corresponding detection statistic is shown in Figure~\ref{fig:detection_statistic_surface} for the surface sensors, together with the background window actually used. In OBSN, and especially in BSDL, short transients are clearly visible, spanning little more than a single or few time bins of the spectrogram ($10$~s resolution). These events are discussed further in Sections~\ref{sec:catalogue} and~\ref{sec:shorts}.

Figure~\ref{fig:waveform_narrowband} shows the waveforms of the four stations on a common 200~s window and the narrowband spectrum computed on the same window with a single Hann-tapered FFT to resolve spectral lines. The narrowband spectrum shows structured lines around $4$, $5$ and $9$ Hz and it can be considered the spectral signature of the event.

\begin{figure}[t!]
\centering
\includegraphics[width=0.7\textwidth]{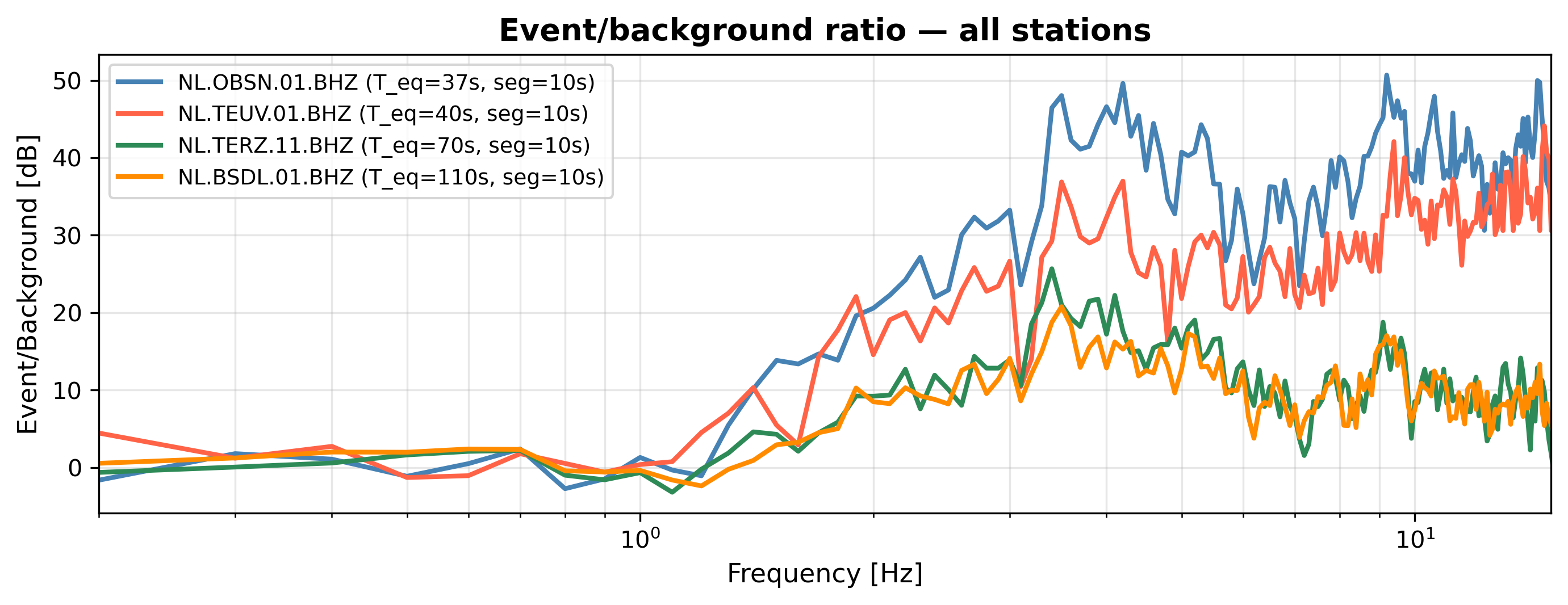}
\caption{Event-to-background PSD ratio for the passage of Figure~\ref{fig:spectrograms_stacked}, at the four borehole sensors. For each station the ratio is computed between the PSD of its own $T_{\rm eq}$ window, centred on the local envelope peak, and the PSD of the background window of the same chunk. The equivalent duration and the Welch segment used are reported in the legend; the growth of $T_{\rm eq}$ from OBSN to BSDL follows the decrease of the signal-to-noise ratio and is discussed in Section~\ref{sec:single_event}.}
\label{fig:event_bg_ratio}
\end{figure}

Figure~\ref{fig:event_bg_ratio} shows the event-to-background ratio at the four stations. The ratio is compatible with 0~dB below 1~Hz and rises steeply above it, reaching 30--50~dB in the 3--15~Hz band, with the largest excess at OBSN, the station closest to the track. The shape of the ratio is remarkably similar at the two stations close to the line, OBSN and TEUV, which differ mainly by an overall offset, whereas TERZ and BSDL depart from it around 10~Hz, where their excess flattens or decreases instead of continuing to rise. This is consistent with the high-frequency content of the transient being attenuated over the propagation distance.

\begin{figure}[!t]
\centering

\begin{subfigure}{\textwidth}
    \centering
    \includegraphics[width=\textwidth]{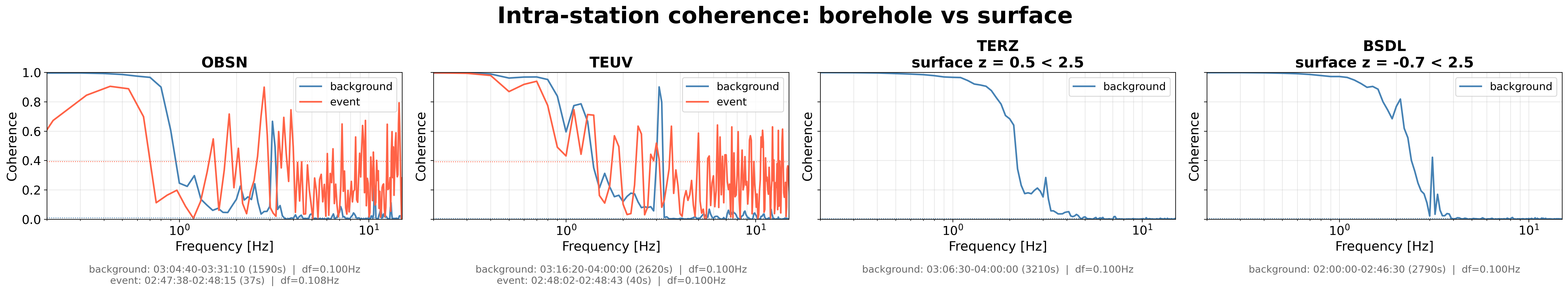}
    \label{fig:coh_intra}
\end{subfigure}

\begin{subfigure}{\textwidth}
    \centering
    \includegraphics[width=\textwidth]{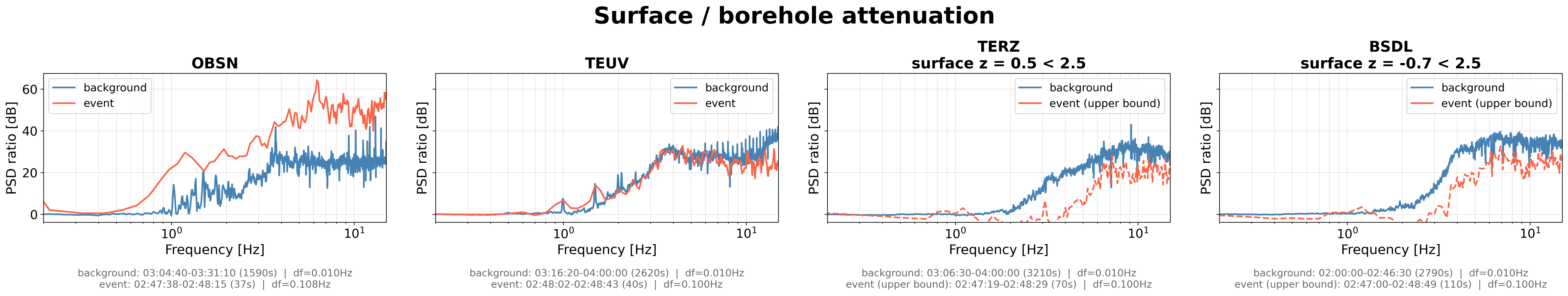}
    \label{fig:atten}
\end{subfigure}

\begin{subfigure}{\textwidth}
    \centering
    \includegraphics[width=\textwidth]{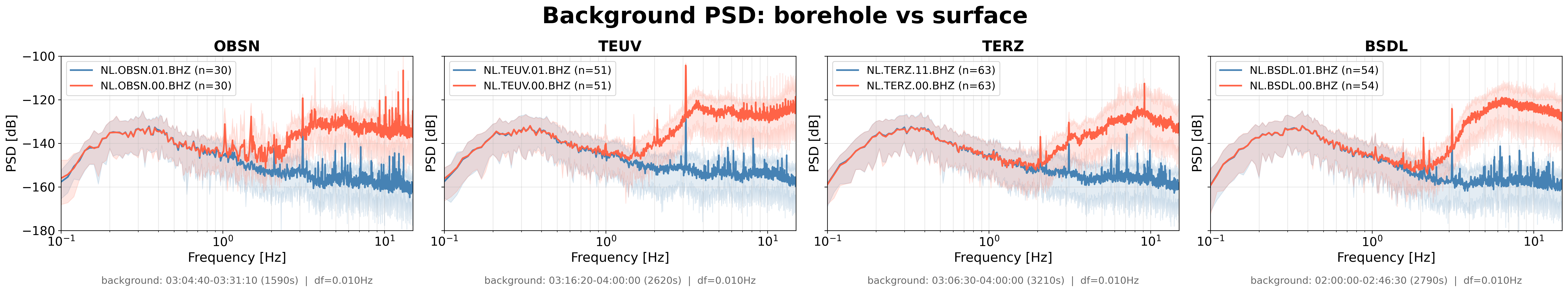}
    \label{fig:psd_bh_sf}
\end{subfigure}
\caption{Each column corresponds to one station and, in every panel, blue refers to the background window and red to the event window. \textit{First row:} surface-to-borehole coherence. The two curves average a very different number of segments and each must be read against its own 95\% significance level (dashed lines). \textit{Second row:} surface-to-borehole PSD ratio, i.e.\ the attenuation provided by the combination of the soft cover and the depth. Where the surface sensor does not cross the detection threshold inside the $T_{\rm eq}$ window, the event curve is replaced by the dashed upper limit given by the surface background. \textit{Third row:} background PSD of the surface and borehole sensors separately, with the shaded bands giving the $5^{\rm th}$--$95^{\rm th}$ percentile range. Welch parameters and the number of averaged segments are reported under each panel.}
\label{fig:event_composite}
\end{figure}

The comparison with the co-located surface sensors is presented in Figure~\ref{fig:event_composite}. In the absence of transients, the surface and the borehole sensor of the same station are almost fully coherent below $\sim 0.5$~Hz, where both are driven by the microseismic field. The background coherence collapses above $\sim 1$~Hz: the ambient high-frequency fields at the surface and at depth are essentially independent. During the event the coherence remains slightly significant, although strongly structured, up to 10~Hz at the stations where the surface crosses the threshold. The two curves average a very different number of segments, a few units for the event against hundreds for the background, and therefore have different statistical floors (dashed lines). 

The surface-to-borehole attenuation grows from $\sim 0$~dB below 0.3~Hz to 40--50~dB above 5~Hz for OBSN. The surface-to-borehole ratio quoted here is not a pure depth attenuation: the two sensors of a station also differ in their distance to the source. At OBSN the surface sensor is ~60 m from the track while the borehole sensor is at 280 m depth, so that geometric spreading alone accounts for a part of the observed ratio, and the remainder mixes anelastic attenuation with the response of the soft cover. The values reported should therefore be read as the total surface-to-borehole rejection of this specific geometry, and not as a depth gain transferable to other configurations.

It should be also noted that the surface background is not stationary, and both the coherence and the attenuation estimated on the event window can be affected by this. At OBSN, the event coherence between surface and borehole decreases below 1~Hz while the background remains fully coherent, and the event attenuation rises at low frequency where the background attenuation is flat at 0~dB. The same behaviour is observed for other transients at OBSN, which points to an effect related to the proximity of the railway line. The surface sensor at TEUV experienced some issues during this event window, resulting in several resonances in the spectra. Consequently, the quantities computed from this sensor should be regarded as indicative.

\subsection{Modelling}

The measurements presented above are intended to be used as input and as validation for the wave-propagation models developed for the EMR site. The modelling chain relies on the TRAFFIC toolbox \cite{traffic}, which computes the ground vibration generated by a train running on a track from the vehicle and track parameters and from the dynamic properties of the layered soil, and which has been validated against dedicated measurement campaigns on test tracks \cite{CARNEIRO2026109768}. Combined with the three-dimensional geological model of the area, this provides a route from an individual convoy to the displacement field at the depth of the boreholes to be compared directly with the event spectra of the catalogue.

A calibrated source model would allow the transient population to be extrapolated to positions where no borehole is installed, in particular to the envisaged corner points, from a known distance and geology. Given the geology of the EMR area, this extrapolation requires the model to be first validated against measurements at or near the corner points, starting with BSDL and then with the boreholes being completed at Hoof and Asse. It would also provide realistic synthetic transients on which the NN subtraction schemes can be trained and tested, since the performance of these schemes on non-stationary signals cannot be assessed on the stationary background alone. Finally, a source model is what makes it possible to explore mitigation options at the level of the source or of the propagation path, which is not accessible from the measurements alone.

In a future work we plan to combine the event catalogue presented here with the timetable and the rolling-stock characteristics of the convoys operating on the line, together with the track and embankment parameters, and to use them as input to TRAFFIC. Comparing the predicted and the observed spectra on an event-by-event basis will allow the modelling chain to be tested against underground sensors at borehole depth.
 
\subsection{The 2026 event catalogue}
\label{sec:catalogue}

\begin{table}[t!]
\centering
\begin{tabular}{lcccc}
\toprule
station & triggers [$z>2.5$] & analyzed hours & rate [h$^{-1}$] & flagged time \\
\midrule
OBSN & $17\,488$ & $5042$ & $3.5$ & $7.5\%$ \\
TEUV & $14\,627$ & $5050$ & $2.9$ & $5.9\%$ \\
TERZ & $2594$    & $3626$ & $0.7$ & $0.5\%$ \\
BSDL & $2493$    & $946$  & $2.6$ & $1.8\%$ \\
\bottomrule
\end{tabular}
\caption{Detection rates over the analysis period. BSDL is restricted to the data after 26 May 2026, as in Section~\ref{sec:background}. Analyzed hours count the time actually covered by valid data and are therefore shorter than the calendar span. The flagged time is the total trigger duration (not the equivalent duration that for the long transients is shorter) divided by the analyzed time. Both the rate and the flagged time depend on the detection parameters and threshold.}
\label{tab:rates}
\end{table}

The detection statistics of the four stations are summarized in Table~\ref{tab:rates}. At this stage the rates are not directly comparable between stations because the catalogues do not contain the same mixture of populations: the median equivalent duration is $5.4\,$s at BSDL against $38$--$70\,$s at the other three stations with OBSN that also carries a substantial number of short-duration triggers. These are of a different nature from the railway transients and are discussed separately in Section~\ref{sec:shorts}. 

\begin{figure}[p!]
\centering
\captionsetup[subfigure]{skip=1pt, justification=centering, singlelinecheck=false, labelfont=bf}
\begin{subfigure}{0.96\textwidth}
    \centering
    \includegraphics[width=\textwidth]{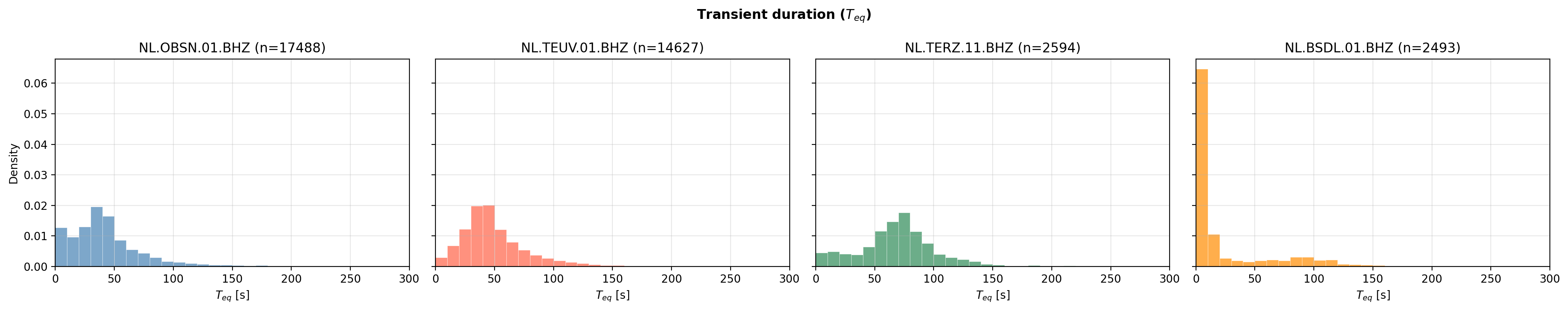}
    \caption{}
    \label{fig:dur_dist}
\end{subfigure}\\[-2pt]
\begin{subfigure}{0.96\textwidth}
    \centering
    \includegraphics[width=\textwidth]{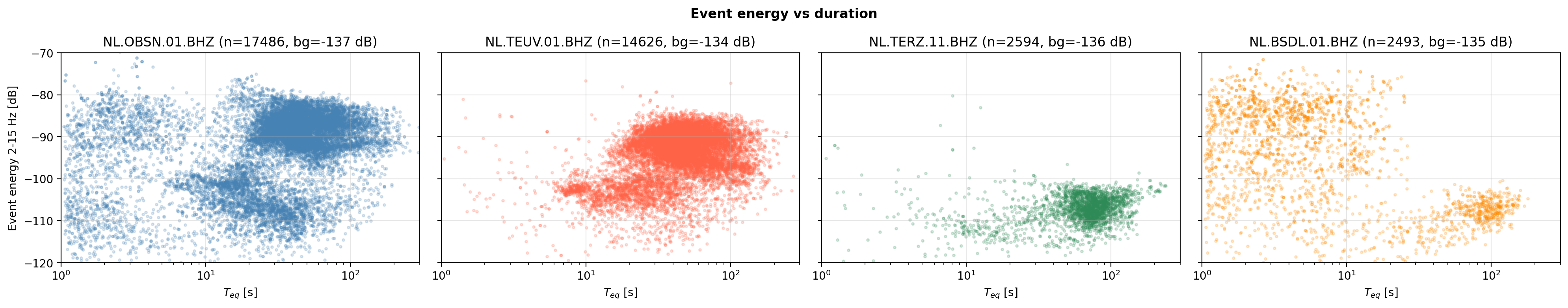}
    \caption{}
    \label{fig:power_dur}
\end{subfigure}\\[-2pt]
\begin{subfigure}{0.96\textwidth}
    \centering
    \includegraphics[width=\textwidth]{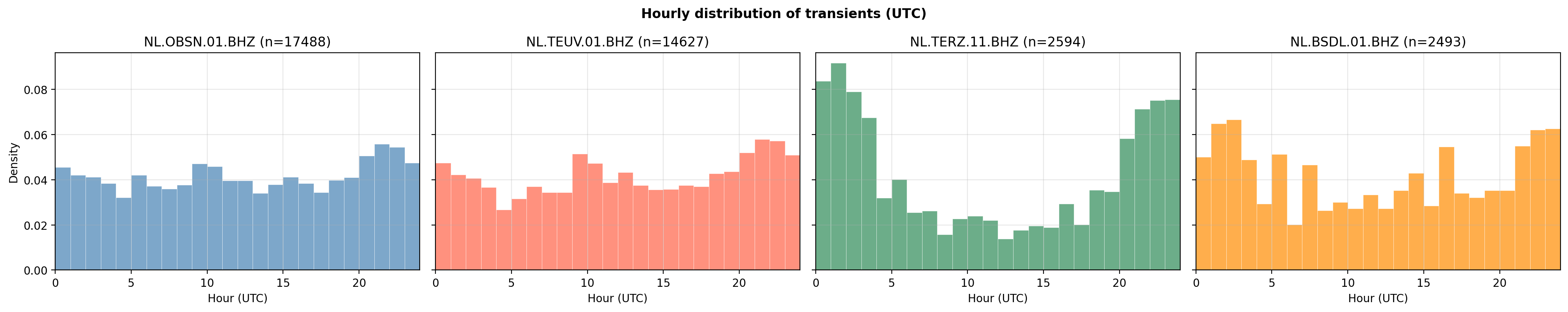}
    \caption{}
    \label{fig:hourly}
\end{subfigure}\\[-2pt]
\begin{subfigure}{0.96\textwidth}
    \centering
    \includegraphics[width=\textwidth]{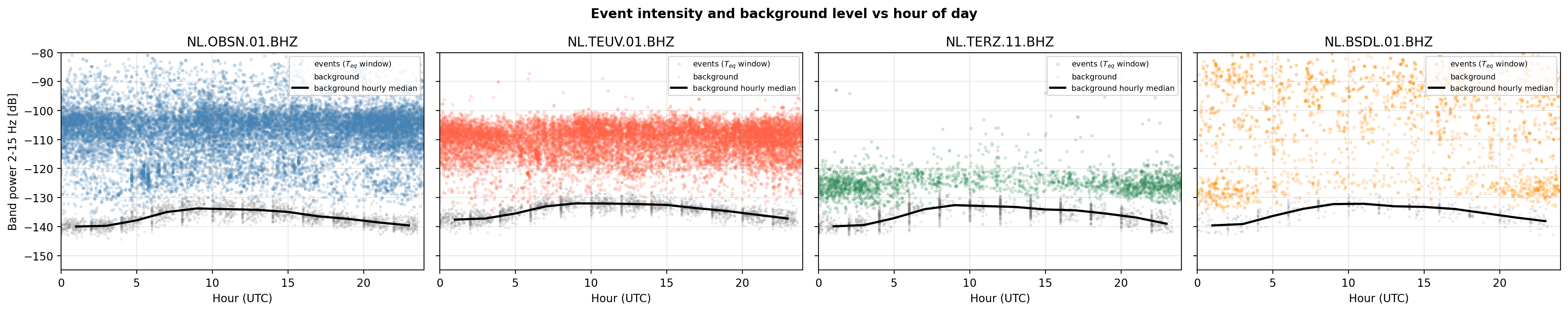}
    \caption{}
    \label{fig:hourlypower}
\end{subfigure}\\[-2pt]
\begin{subfigure}{0.96\textwidth}
    \centering
    \includegraphics[width=\textwidth]{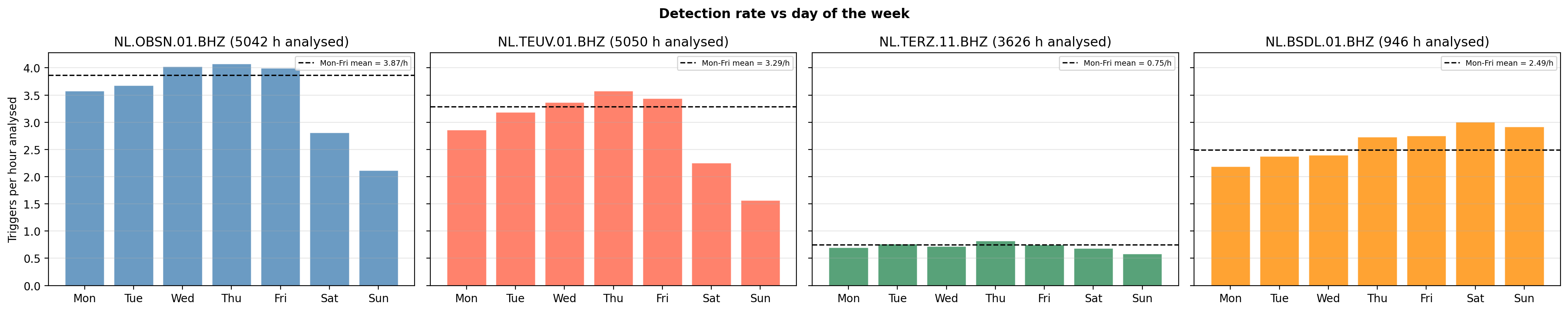}
    \caption{}
    \label{fig:weekday}
\end{subfigure}
\caption{Statistical characterization of the transients detected at the four stations over the first seven months of 2026, with BSDL restricted to the data after 26 May. \textit{First row:} distribution of the equivalent duration $T_{\rm eq}$ defined in Section~\ref{sec:zscore}. \textit{Second row:} $2$--$15\,$Hz energy of each event, as a function of $T_{\rm eq}$; a population radiating a constant energy would be flat in this plane and one radiating a constant power would rise by $10\,$dB per decade. The median band power of the background windows is quoted in each panel title as a reference level. \textit{Third row:} hourly distribution of the detections. \textit{Fourth row:} absolute band power of the events (coloured) and of the background windows (grey), with the hourly median of the background; the comparison shows that the hourly patterns of the third row are partly a modulation of the detection threshold and not of the sources. \textit{Fifth row:} detection rate as a function of the day of the week, normalized to the hours actually analyzed for each weekday; the dashed line is the Monday--Friday mean.}
\label{fig:trigger_statistics}
\end{figure}

First row of Figure~\ref{fig:trigger_statistics} shows the distribution of the equivalent durations $T_{\rm eq}$ defined in Section~\ref{sec:zscore}. At OBSN, TEUV and TERZ the distributions are unimodal and positively skewed, with medians of $37.6$, $43.6$ and $70.0\,$s respectively. BSDL (as also OBSN) is qualitatively different, its distribution piling up at the shortest durations with a median of $5.4\,$s, for the reason discussed above. The larger median at TERZ should be read together with the behaviour of the estimator described in Section~\ref{sec:single_event}: $T_{\rm eq}$ grows as the signal-to-noise ratio decreases, and TERZ is the station with the lowest one. 

Second row shows the energy in the $2-15$Hz band of each event as a function of the duration. The energy measures the total disturbance deposited at the site: a long weak transient and a short loud one can perturb it by the same amount. No clear trend with duration is present at any station but the plane shows that the catalogue is not a single population. At OBSN the events separate into two branches about $12\,$dB apart in energy, with a gap between them that a threshold effect alone would not produce, and a similar although less pronounced structure is visible at TEUV. This separation in energy is the starting point of the classification of Section~\ref{sec:coincidence}.

Third row shows the hourly distribution of the detections. OBSN and TEUV, the two stations closest to the railway, show an almost flat pattern. TERZ shows a marked pattern with maxima in the early morning and late evening and a pronounced minimum during working hours, while BSDL shows the opposite behaviour, with more energetic events during the day.

These three patterns have different origins and it is important not to read them all as a modulation of the source. Indeed, the fourth row shows the absolute band power of the detected events, measured over their $T_{\rm eq}$ window, together with that of the background windows, on the same scale. The event level is essentially independent of the hour of the day, whereas the background follows a clear diurnal cycle of $5$--$8\,$dB, with the minimum around 02--04 UTC and the maximum around 10--12 UTC, at all four stations. Since the detection threshold is defined relative to that background, the same physical disturbance is more easily detected at night, and the nocturnal excess of detections at TERZ is largely a selection effect of the relative threshold. The flat pattern of OBSN and TEUV, where events exceed the local floor by tens of dB, is instead robust against this bias.

Fifth row shows the detection rate as a function of the day of the week, normalized to the hours actually analyzed for each weekday, so that the uneven coverage of the four stations does not enter the comparison. A population dominated by the freight train traffic is expected to show a moderate reduction at the weekend.

The panels of Figure~\ref{fig:trigger_statistics} describe the catalogue as a single population and show that it is not one. However, assigning each event to a source is not possible with the present data: no reliable database of the railway traffic is available for the period, and none of the observables shown here is by itself decisive on a single event. What can be done, and is the subject of the following section, is to use the coincidence between stations as a statistical ordering criterion, and to describe the populations that emerge from it.

\subsection{Coincidence and tentative classification}
\label{sec:coincidence}

\begin{figure}[t!]
\centering
\includegraphics[width=0.8\textwidth]{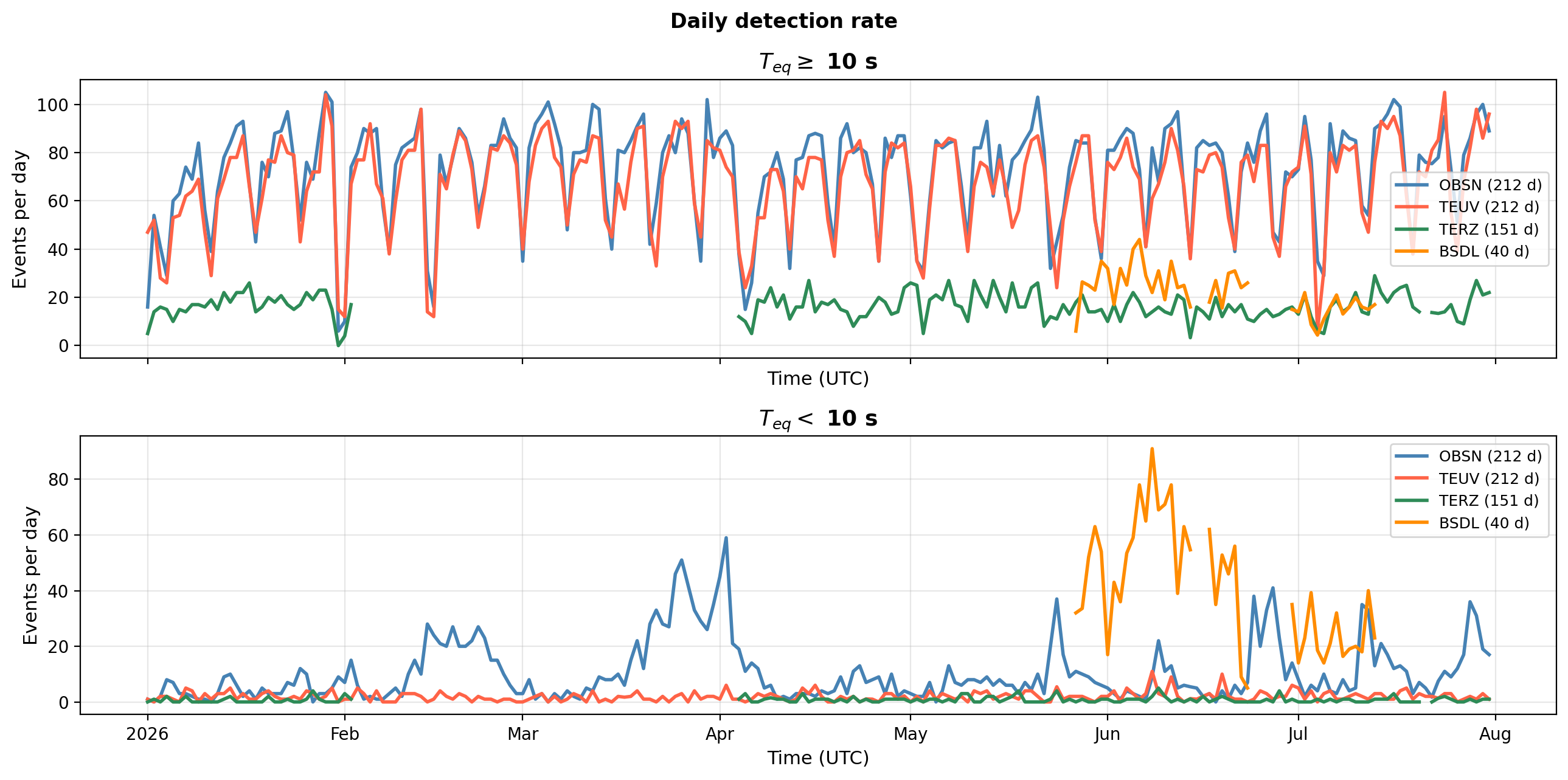}
\caption{Daily detection rate at the four stations, split at $T_{\rm eq}=10\,$s. \textit{Top:} long triggers, showing a stable weekly cycle at the stations closest to the train line. \textit{Bottom:} short triggers, discussed in Section~\ref{sec:shorts}, which occur in isolated episodes, concentrated in time and not shared between stations.}
\label{fig:num_over_time}
\end{figure}

The aim of this section is not to label individual events but to look for structure in the catalogue, using the coincidence between stations as an ordering criterion. BSDL is not used, since its period of simultaneous coverage with the other stations is too short. Figure~\ref{fig:num_over_time} shows the daily detection rate split at $T_{\rm eq}=10$~s: the long triggers follow a stable weekly cycle at OBSN and TEUV over the whole period, while the short ones appear in isolated episodes, concentrated in time and not shared between stations; they will be discussed in Section~\ref{sec:shorts}.

Two triggers at different stations are flagged as "coincident" when their z-score peaks are within $|\Delta t|\le W=100$~s. The peak marks the closest approach of the source and depends less on the signal-to-noise ratio than the trigger onset. Pairs are formed one-to-one, starting from the smallest $|\Delta t|$, and delays are quantized at 10~s by the spectrogram grid of the detector. Since the dominant source is a moving train, the delay is a transit delay, i.e. the along-track separation of the two closest-approach points divided by the speed, and its sign reverses with the direction of travel.  Triggers within 25~s of the predicted first arrival (with \texttt{iasp91} model \cite{iasp91}) of a catalogue event are labelled \emph{earthquake} and are exempt from this correction. The catalogue merges the FDSN services of EMSC, KNMI, ISC and RESIF without magnitude floor: within 120~km it contains 72 unique events. Of these, 27 are detected at OBSN, 24 at TEUV, 21 at TERZ; as discussed in Appendix~\ref{app:earthquakes}, teleseisms do not enter the statistic. Each pair of stations is analysed only on the days in which both have at least six chunks with a background estimate, i.e.\ 212 days for TEUV--OBSN and 151 when TERZ is included. The spectral shape is the median, over its events, of the ratio between the event PSD, computed over a window of length $T_{\rm eq}$ centred on the envelope peak, and the background PSD of the nearest quiet window within one chunk (computed with the same 10 s Welch segments as for the event PSD).

The number of accidental pairs is estimated with time slides of one station by whole weeks, which give the probability that two unrelated triggers fall within $W$. Since the pairing is one-to-one, a trigger already matched to its own passage cannot also form a random pair, so this probability is applied only to the triggers left without a genuine partner, $N_{acc}\simeq r_a r_b\,2W\,T$, where $r_a$ and $r_b$ are the rates of unmatched triggers at the two stations and $T$ is their common live time. This gives $128\pm11$ accidental pairs out of 1865 for TERZ--OBSN and $140\pm12$ out of 12335 for TEUV--OBSN. 

\begin{figure}[t!]
\centering
\includegraphics[width=\textwidth]{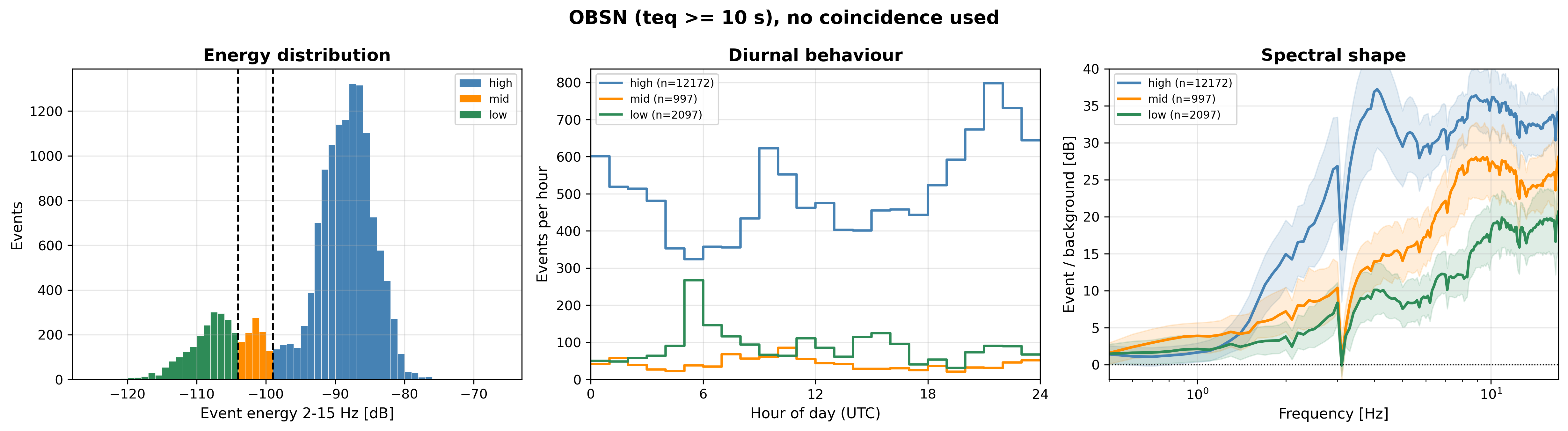}
\caption{Events with $T_{\rm eq}\ge 10\,$s at OBSN, classified by their energy between 2 and 15~Hz; no coincidence information is used. \textit{Left:} energy distribution and the two cuts separating the three groups. \textit{Centre:} diurnal distribution of the three classes. \textit{Right:} spectral shape, with the median and interquartile range of each class.}
\label{fig:obsn_classification}
\end{figure}

OBSN is the station closest to the line with the highest signal-to-noise ratio, and it is the only one in which populations can be separated without coincidence. For the 15266 triggers with $T_{\rm eq}\ge 10$~s, the event energy in 2--15~Hz shows three groups separated by two cuts (Figure~\ref{fig:obsn_classification}). The high-energy class dominates the catalogue, has a nearly flat diurnal distribution and a spectral peak at $\sim 4$~Hz, and is the one most plausibly associated with the trains. The mid class has a spectral shape rising with frequency and a mild excess during the day. The low class is concentrated around 06~UTC, which suggests a fixed installation. These associations are not established: the classes are defined by the data and their interpretation remains tentative.

\begin{figure}[t!]
\centering
\includegraphics[width=\textwidth]{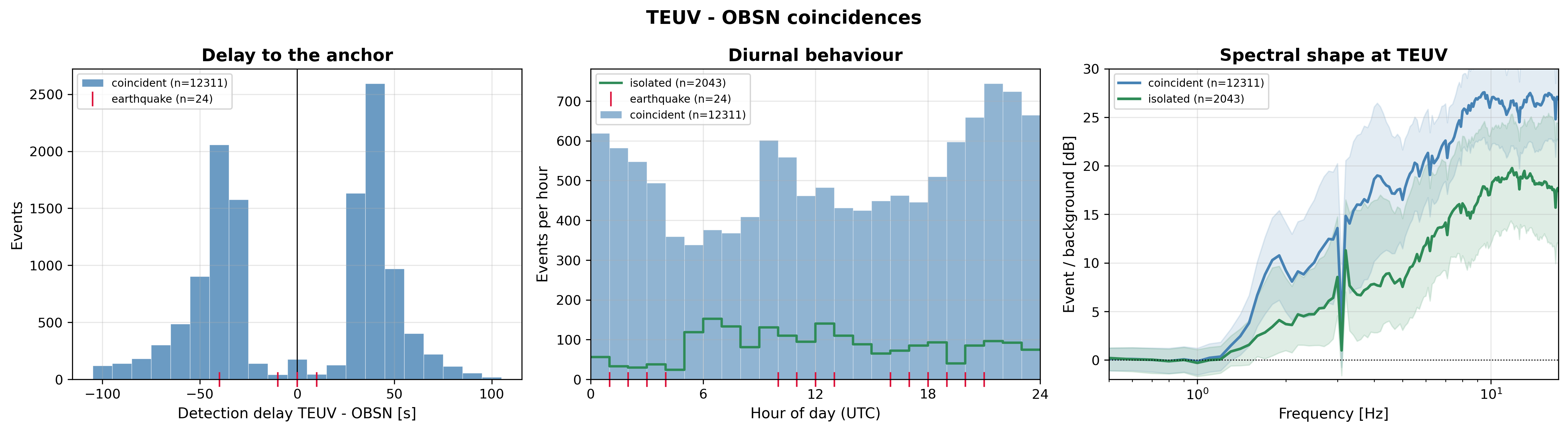}
\caption{Transients at TEUV, classified by coincidence with OBSN. \textit{Left:} delay between the peaks at TEUV and at OBSN; the maxima at $\pm 40\,$s correspond to the two directions of travel. Red ticks mark earthquakes. \textit{Centre:} diurnal distribution of the coincident, isolated and earthquake classes. \textit{Right:} spectral shape of the coincident and isolated classes.}
\label{fig:teuv_classification}
\end{figure}

At TEUV the energy distribution does not separate into groups and the coincidence with OBSN is used instead. Of the detected triggers on valid days, the 84\% are coincident. The delay distribution (Figure~\ref{fig:teuv_classification}) has two sharp maxima at $\pm 40$~s, the signature of trains travelling in the two directions. With an along-track separation of $\approx 0.9$~km between the two closest-approach points and the 10~s quantisation, this corresponds to 65--110~km/h. Earthquakes fall at $\Delta t\simeq 0$, as expected for a wave crossing 1~km. The isolated events are about 9~dB weaker and are more concentrated during the day; their origin is not established.

\begin{figure}[t!]
\centering
\includegraphics[width=\textwidth]{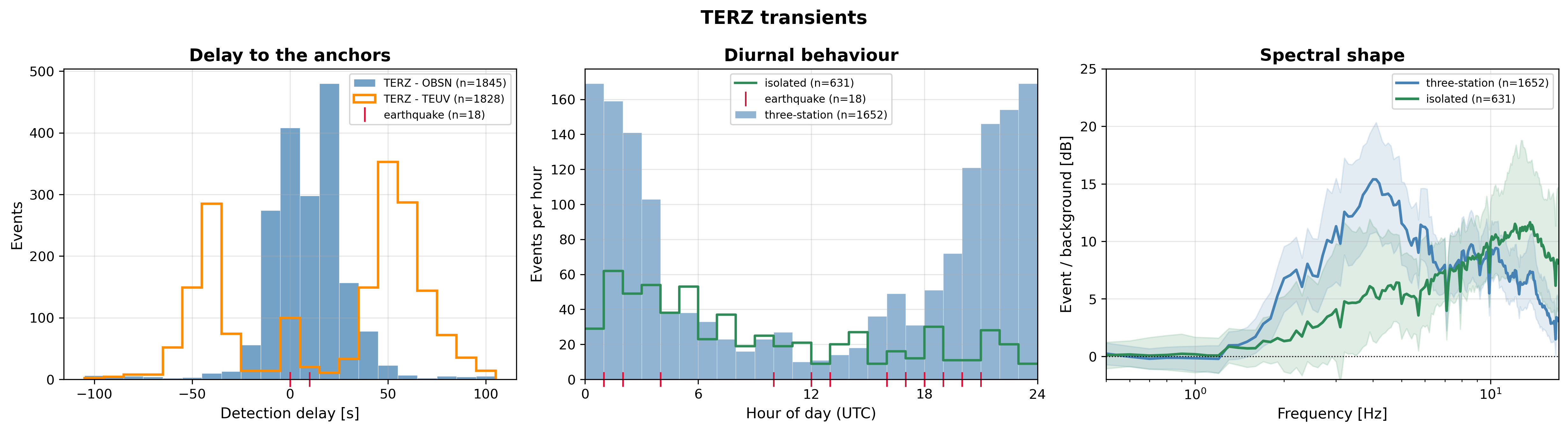}
\caption{Transients at TERZ. \textit{Left:} delays with respect to OBSN and TEUV for the triggers coincident with at least one of them. \textit{Centre:} diurnal distribution of the three-station coincidences with $|\Delta t|\ge 15\,$s with respect to TEUV, of the isolated events and of the earthquakes. \textit{Right:} spectral shape of the three-station coincidences and of the isolated events.}
\label{fig:terz_classification}
\end{figure}

TERZ, the station with the fewest triggers, is used as the key in Figure~\ref{fig:terz_classification}. A TERZ trigger is coincident if OBSN or TEUV has a peak within $\pm 100$~s, and isolated otherwise. The delays to TEUV keep the bimodal transit structure, while those to OBSN cluster near zero, at about $-10$ and $+20$~s depending on the direction. TERZ therefore peaks at an along-track point a few hundred metres beyond OBSN, on the side opposite to TEUV, compatible with the foot of its perpendicular (close to the Remersdaal viaduct). The delays to TEUV also show a narrow excess at zero, which cannot be a transit. Of the 83 three-station coincidences with $|\Delta t|<15$~s with respect to TEUV, 69 are at zero delay with OBSN as well. They are simultaneous detections at three sites up to 2~km apart, with similar amplitude and tail of 15--20~s. They are concentrated between 10 and 13~UTC, which points to quarry blasts. The centre and right panels use the three-station coincidences with $|\Delta t|\ge 15$~s with respect to TEUV. These are detected mostly at night, only 18\% between 06 and 18~UTC because at $\sim 2$~km the train emerges above the relative threshold only when the noise is low. Their spectral shape is consistent with the loss of high frequencies along the path shown below. The isolated events at TERZ peak at higher frequency and are not further classified.

\begin{figure}[t!]
\centering
\includegraphics[width=\textwidth]{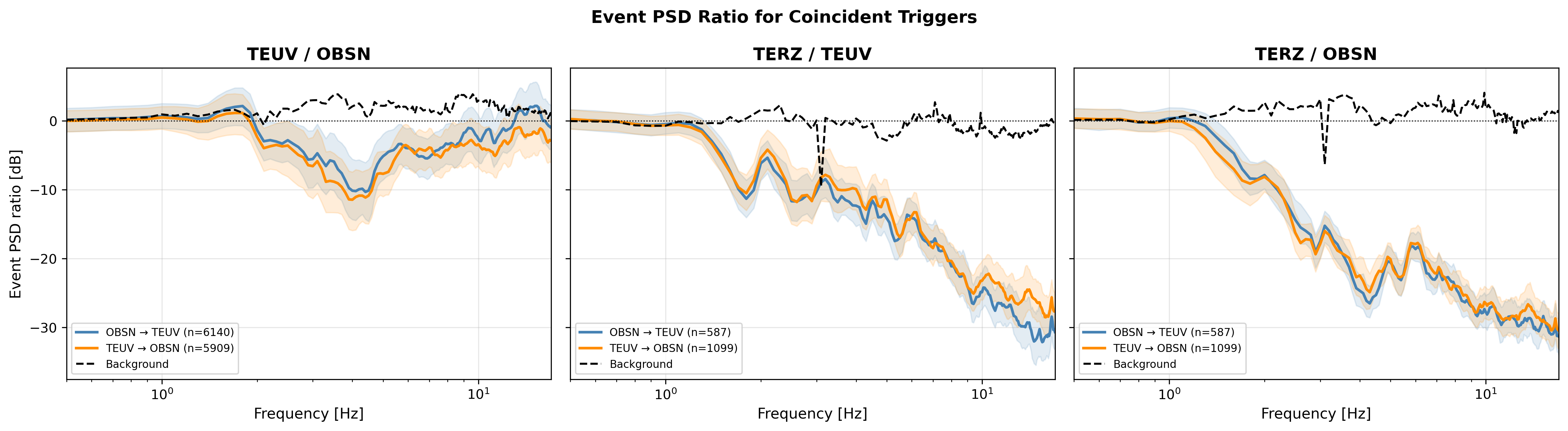}
\caption{Median ratio between the event PSDs of the same train at two stations, with the band between the 25th and 75th percentiles, for the TEUV--OBSN coincidences with $|\Delta t|\ge 15\,$s. \textit{Left:} TEUV/OBSN. \textit{Centre:} TERZ/TEUV, for the TERZ triggers coincident with the TEUV trigger of one of these trains. \textit{Right:} TERZ/OBSN, for the same TERZ triggers and the OBSN trigger of the same train. In all panels the colour gives the direction of travel, from the sign of $t_{\rm TEUV}-t_{\rm OBSN}$, and the dashed line is the ratio of the corresponding background PSDs.}

\label{fig:psd_ratios}
\end{figure}

Figure~\ref{fig:psd_ratios} compares the stations on the same transient. The two PSDs are measured at different times, each at the closest approach to its own station, so the ratios reflect the combined response of site and path. The direction of travel is taken in all panels from the sign of $t_{\rm TEUV}-t_{\rm OBSN}$. Below $\sim 1.5$~Hz the event ratio coincides with the background ratio, because the train does not emerge from the noise. Above 8~Hz TEUV/OBSN is within 1--3~dB of zero, as expected from the similar sensor-to-track distances (289 and 341~m including depth), while between 3 and 5~Hz it drops to $-8$/$-9$~dB. The ratio alone does not tell whether this is an excess at OBSN, 60~m from a railway bridge and with its high-energy class peaking at 4~Hz, or a deficit at TEUV, where the line runs in a tunnel. The ratios involving TERZ decrease steadily with frequency, down to $\approx -25$~dB with respect to TEUV at 8--15~Hz, with a background ratio close to 0~dB, consistent with a strong frequency-dependent loss over the $\sim 2$~km path to TERZ. The spectral features do not depend on the direction of travel, but the detection rate at TERZ does: the two directions are balanced at TEUV and OBSN (5909 and 6140 trains), whereas TERZ detects 1099 trains moving towards OBSN and only 587 in the opposite direction. The former are recorded at OBSN with $\sim 2.8$~dB more power between 3 and 8~Hz and have a longer $T_{\rm eq}$ (median 46~s against 38~s). Since the path loss brings the TERZ detections close to the threshold, a source difference of this size is enough to change the fraction of trains that emerge there, and the longer $T_{\rm eq}$ points to slower or longer convoys in that direction. A lower detection threshold at TERZ would likely recover part of the missing transits, at the cost of more noise triggers and accidental coincidences.

\subsection{Short transients}\label{sec:shorts}

The triggers with $T_{\rm eq}<10\,$s at OBSN and BSDL form a population with properties opposite to those discussed so far, and are treated separately. They appear in isolated episodes showing no diurnal or weekly modulation as visible in the bottom panel of Figure~\ref{fig:num_over_time}. They also have no coincident trigger at the co-located surface sensors.

On the raw counts they are single impulses of 0.1--0.5~s (showing the limits of the $T_{\rm eq}$ definition for the shortest transients), reaching hundreds of times the local noise level, with no scattering tail. The absence of a tail, together with the very short rise time, suggests that a substantial fraction of their energy lies above 15~Hz, so that the durations reported here are set by the analysis band. A 
test at BSDL shows them also on the horizontal component, with a median vertical signal-to-noise ratio of 36~dB. The comparison between the horizontal and the vertical components of the typical events at OBSN is more informative: the horizontal components are affected considerably more than the vertical one, and while the vertical component shows only a short transient, the horizontal components creep slowly back to their original level. This is likely the signature of a tilt of the sensor, subsequently compensated by the instrument, and therefore of a minuscule physical movement of the sensor.

No conclusion is drawn on the nature of these triggers. Their properties are compatible with a local mechanical origin at the borehole itself, whether of the installation or of the instrument. A dedicated analysis is needed to establish whether they are instrumental glitches; until then they have been excluded from the catalogue analysis of Sections~\ref{sec:catalogue} and \ref{sec:coincidence}.

\section{Discussion}
\label{sec:conclusion}

We have presented the first results of the seismic characterization of the EMR candidate site based on the borehole network recently installed in the Terziet area, covering the first seven months of 2026 at four underground stations together with their co-located surface sensors. We also introduce \texttt{SiteChar}, a new Python-framework, which provides a common processing pipeline that allows to separate the stationary background from the transient contributions.

The results have to be read in the light of how the network was deployed. Two of these boreholes were deliberately positioned next to railway infrastructure and only BSDL lies close to a possible corner point, so what the network measures is how anthropogenic noise propagate in depth at different distances from the sources. 

The central methodological result is the separation between the stationary background and the transient component, made necessary by the fact that the field is non-stationary on several timescales. Once the fast non-stationarities are excluded, the four boreholes, drilled in similar geological conditions and at distances from the railway ranging from tens of metres up to about two kilometres, describe essentially the same stationary field. The slower non-stationarities survive in the background spectra themselves, which retain a diurnal cycle of $5$--$8$~dB above 1~Hz and the seasonal modulation of the microseism.

The transient population is dominated by railway traffic and is detected at all four stations, including TERZ and BSDL, more than two kilometres from the line and at hundreds of meters underground. The coincidence between stations resolves the transit of the source along the line, giving an average speed consistent with the line speed, and separates the catalogue into classes that differ in energy, in diurnal distribution and in spectral shape. During a passage the borehole and the surface sensor of the same station remain coherent above the frequency at which the ambient fields decorrelate, so that a surface sensor carries usable information on the disturbance seen at depth. At the stations far from the line the passage stays above the borehole noise floor while falling below the surface one, which quantifies event by event the gain in detectability provided by the depth. 

These transients are not a marginal correction to the ambient field: they exceed the local background by tens of dB in the band where the detector is most closely coupled to its environment, and they occur a few times per hour at the stations close to the line. 

Although this work focuses on vertical seismic measurements, the observed transients already exhibit strong variations across the network, demonstrating that the seismic field is spatially heterogeneous and that disturbances do not affect all underground locations equally. This is relevant for NN modelling because such spatial variability challenges the assumptions of a diffuse, isotropic seismic field underlying analytical NN estimates. Instead, those transients should be modeled based on the sources and wave propagation in the local geology. However, we do not claim that the vertical spectra presented here are sufficient to provide a reliable stand-alone estimate of the NN. Rather, they provide empirical constraints on the seismic environment that can guide realistic numerical modeling, which must account for both horizontal and vertical motion and their spatial correlations. Future work will constrain all these ingredients sufficiently to support a robust site-specific NN prediction, which is needed to go beyond the preliminary estimates presented in the past.

An important consequence is that we deliberately refrain from assigning a premature NN estimate to the seismic characterization presented in this paper. The realistic conversion from the measured seismic field to NN is not a simple analytical rescaling of a seismic noise PSD.
At the single-mirror level, the NN acting on each individual test mass is determined by the surrounding three-dimensional seismic field and its coupling to gravitational-field fluctuations at the test-mass location. 
At the single ET-vertex level, a relevant quantity is the correlated NN acting on the four test masses sharing one corner of the envisioned triangular design. This correlated NN is linked to the spatial seismic coherence between the nearby test masses, both within one interferometer (the two input test masses at a single vertex) and between the input and end test masses of the three different interferometers. These correlations cannot be reliably estimated from the PSD of a single measurement point, and ideally require measurements underground over distances of a few hundred meters. Measurements between boreholes on the kilometer scale, as used in this work, are not reliable estimates of these local correlations, as seismic coherence falls rapidly with distance, but they are still useful for establishing boundary limits.
At the single-interferometer level, the contributions from all three contributing vertices, independent of the final ET layout, must be propagated into the differential arm-length channel of one Michelson interferometer, accounting for their potential correlations.
As this study considers data from one corner point, we stress again that these successive levels require information beyond the local seismic spectra presented in this work. This information will be provided based on available data in future work, with the aim of developing a reliable and detailed NN model tailored to the envisioned ET geometry in the EMR region.

Once such a NN model is established, it will also provide key insights for improving the intended NN mitigation systems. These systems need to differentiate between the seismic background and transient contributions, and their performance will be significantly affected by the seismic wavefield composition and the local seismic coherence. Supplying the transient events, together with the background levels on which the propagation and NN models are built, is the contribution of this work to the integrated modelling chain currently developed for the EMR candidate site.

\section*{Data availability}

Data for the considered boreholes are part of the Netherlands Seismic and Acoustic Network \cite{KNMI} and are publicly available through \url{https://rdsa.knmi.nl/network/NL/}. 

\section*{Acknowledgements}

This work was carried out as part of the Einstein Telescope site-feasibility studies in the Euregio Meuse–Rhine (ET-EMR), coordinated by the ET-EMR Project Office. These studies are conducted within the framework of the Declaration of Intent on regional cooperation for the Einstein Telescope, signed by the responsible ministers of the participating regions and countries.
The authors gratefully acknowledge the Royal Netherlands Meteorological Institute (KNMI) for its responsibility for the operation and maintenance of the seismic monitoring network. We thank the landowners, municipalities and regional authorities in the Euregio Meuse–Rhine for facilitating access to the measurement locations, as well as the ET-EMR partners for their scientific, technical and organisational contributions.The ET-EMR site-feasibility studies received financial support from the Dutch National Growth Fund, the Dutch Ministry of Education, Culture and Science, the Dutch Province of Limburg, the Research Foundation - Flanders (FWO), the Public Service of Wallonia (SPW), the Belgian Fund for Scientific Research (FNRS), the Fédération Wallonie-Bruxelles (FWB), the Belgian Science Policy Office (BELSPO), and the State of North Rhine-Westphalia (NRW).

The authors thank all the members of the EMR noise group for the many discussions on the results presented in this paper. 

Claude Opus 5 (Anthropic) was used for code refactoring and script formatting. The final code and analysis were thoroughly tested and verified by the authors, who accept full responsibility for the software and results presented.

\appendix

\section{Microseism}\label{app:microseism}

\begin{figure}[t!]
\centering
\includegraphics[width=0.75\textwidth]{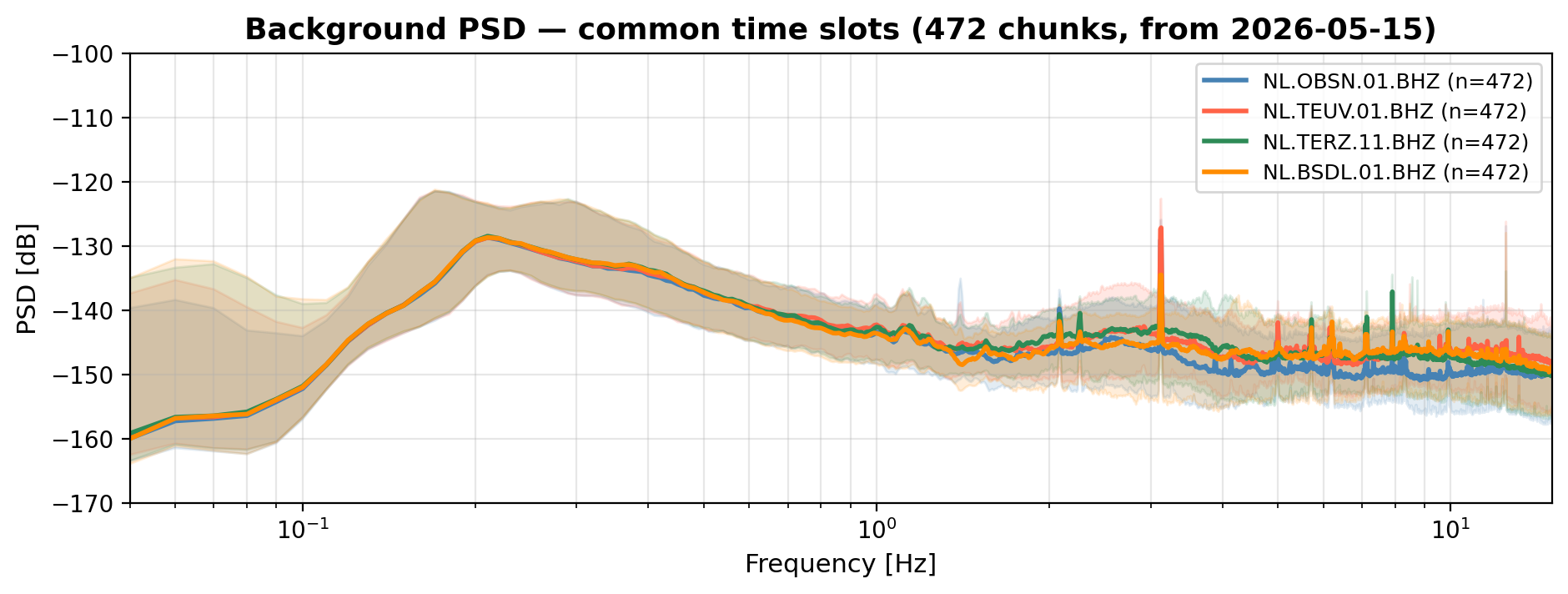}
\caption{PSD on common chuncks between TERZ, TEUV, OBSN and BSDL confirming that the microseism is the same when considering the same time periods.}
\label{fig:common_chunks}
\end{figure}

Figure~\ref{fig:psd_overlay} shows that the secondary microseismic peak at $\sim 0.2$~Hz dominates the spectra of all four stations below $\sim 0.5$~Hz, with median levels between $-124$ and $-117$~dB. The spread of a few dB between stations seen in that figure is not a site property but a consequence of the different temporal coverage of the four datasets, since the microseismic amplitude is modulated by the seasonal cycle of ocean-wave activity. Restricting the comparison to the two-hour chunks that are simultaneously available at OBSN, TEUV, TERZ and BSDL (472 chunks) removes the effect. Figure~\ref{fig:common_chunks} shows that the microseismic peaks then coincide to within a fraction of a dB, both in level and in the width of the percentile band. At these frequencies the field is therefore regionally coherent and identical across
the EMR area, as expected. 

\begin{figure}[t]
\centering
\includegraphics[width=\textwidth]{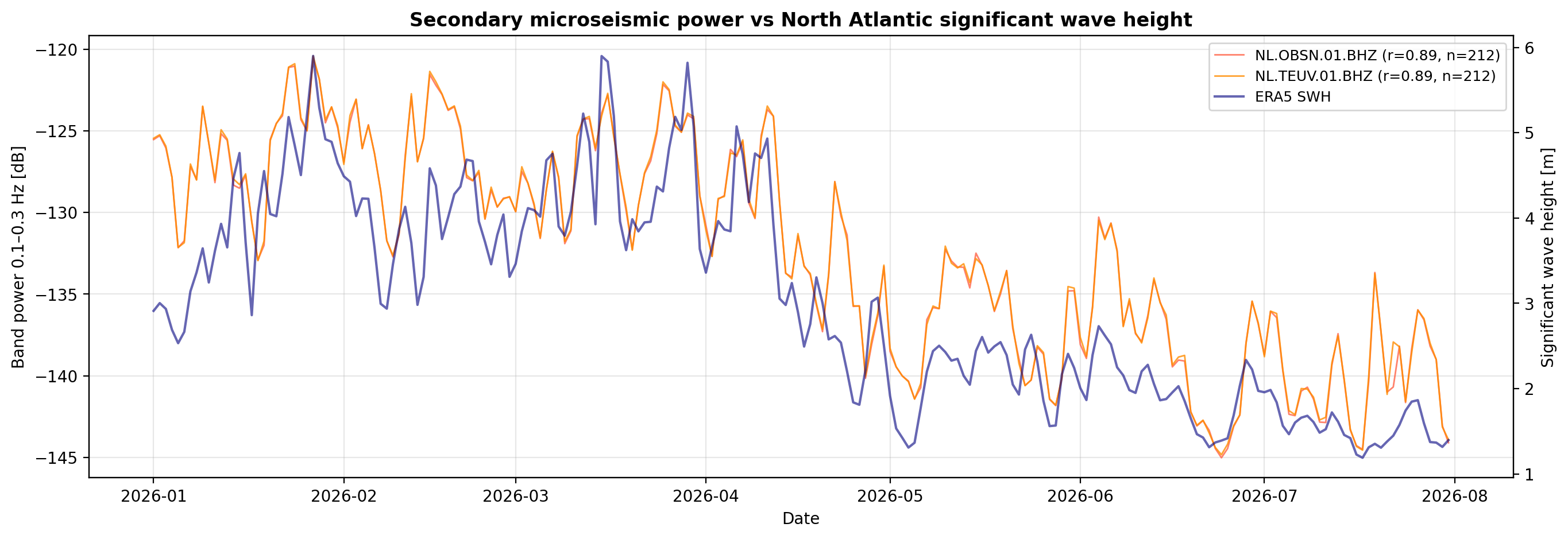}
\caption{Daily median microseismic band power (0.1--0.3\,Hz, background windows) at OBSN and TEUV compared with the area-averaged significant wave height (SWH) over the North Atlantic (40--65$^\circ$N, 40$^\circ$W--5$^\circ$E) from ERA5 reanalysis. Both seismic stations track the ocean wave forcing closely, with Pearson correlation coefficients reported in the legend. The strong positive correlation confirms that the secondary microseismic peak observed at the ET candidate site is driven by North Atlantic storm activity, consistent with the double-frequency generation mechanism. The dual $y$-axis shows seismic power in dB (left) and SWH in metres (right).}
\label{fig:microseism_swh}
\end{figure}

The origin of the modulation is confirmed by the comparison with the ocean forcing. Figure~\ref{fig:microseism_swh} shows the daily median band power in 0.1--0.3~Hz, computed on the background windows only, together with the significant wave height (SWH) area-averaged over the North Atlantic ($40$--$65^\circ$N, $40^\circ$W--$5^\circ$E) from the ERA5 data of Copernicus \cite{era5}. The two series track each other closely over the whole observation period, including the individual storm episodes of January--March and the decay towards May--June, with a Pearson correlation coefficient $r = 0.89$ at both OBSN and TEUV. This is the behaviour expected from the double-frequency generation mechanism, in which the seismic power scales with the interaction of opposing wave trains in the source region.

\section{Earthquakes}\label{app:earthquakes}

Earthquakes appear in the time evolution of the background spectra (Figure~\ref{fig:heatmap_bg_psd}) as coincident horizontal stripes, i.e.\ as short-lived excursions that are broadband in frequency and simultaneous at all stations. They are a natural population of transients, distinct from the anthropogenic one that is the subject of Section~\ref{sec:transients}, and it is useful to establish how they enter the detection statistic.

\begin{figure}[t!]
\centering
\includegraphics[width=0.7\textwidth]{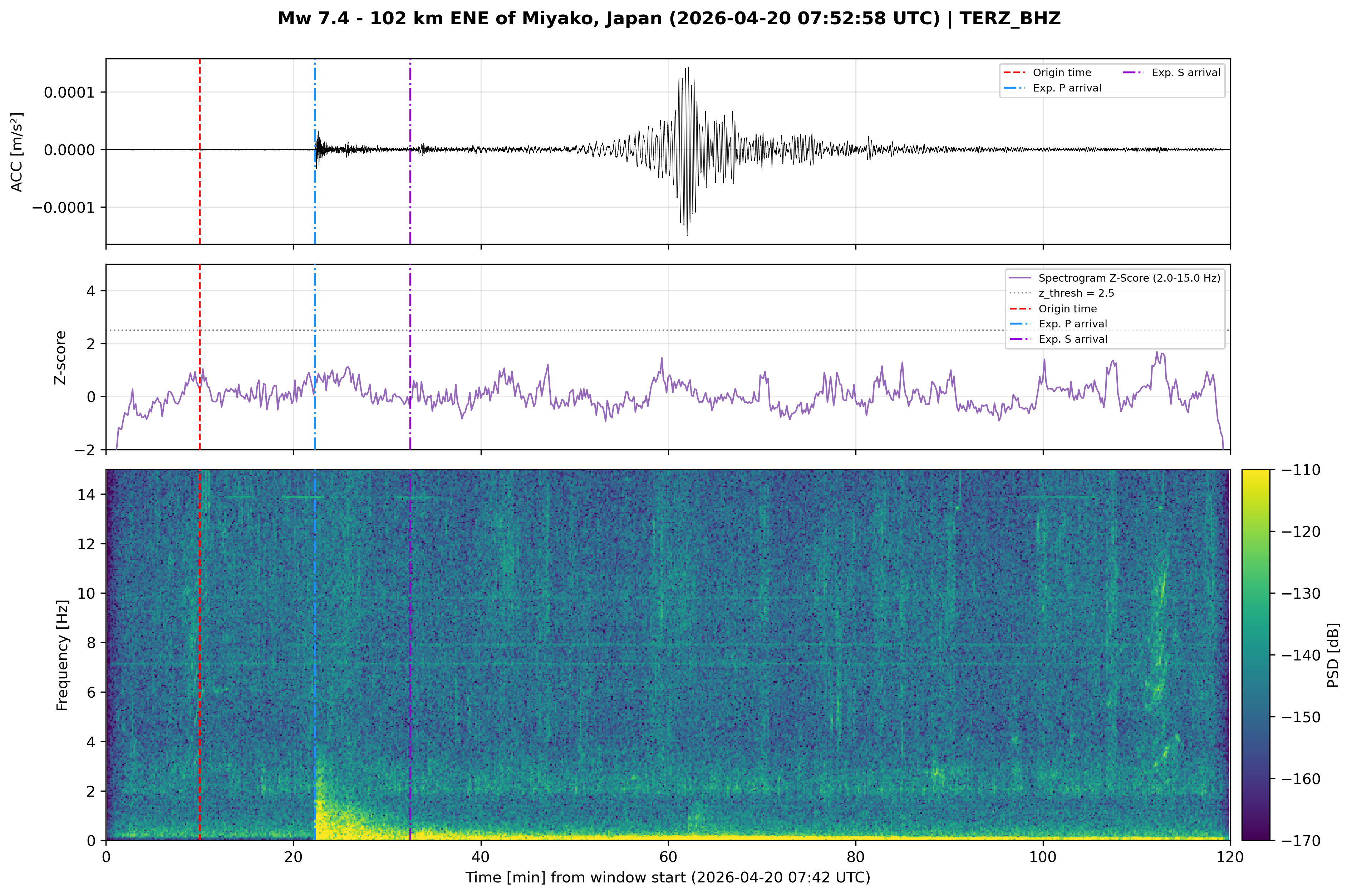}
\caption{Seismic signal analysis of the $M_w\ 7.4$ Miyako, Japan earthquake recorded at station TERZ. 
\textit{Top panel:} Instrument-corrected acceleration waveform ($\text{m/s}^2$). 
 \textit{Middle panel:} Spectrogram Z-score time profile computed in the $[2.0, 15.0]\text{ Hz}$ frequency band; the horizontal dotted line marks the detection threshold ($z_{\text{thr}} = 2.5$). 
\textit{Bottom panel:} Power Spectral Density (PSD) spectrogram in dB. 
In all panels, the red dashed line denotes the event Origin Time (2026-04-20 07:52:58 UTC), while the blue and purple dash-dotted lines indicate the theoretical $P$-wave and $S$-wave arrival times computed via the \texttt{iasp91} model in \texttt{obspy}.}
\label{fig:earthquake}
\end{figure}

Figure~\ref{fig:earthquake} shows a representative case, the $M_w$ 7.4 Miyako (Japan) earthquake of 2026-04-20 recorded at TERZ: the instrument-corrected acceleration waveform shows the arrivals at the predicted times, and the wave train is clearly visible in the low-frequency part of the spectrogram.
The trigger detection follows the same time-frequency $z$-score methodology developed in Section~\ref{sec:zscore}, applied to the same 2--15~Hz band. Theoretical body-wave travel times $t_P$ and $t_S$ are computed with the \texttt{iasp91} \cite{iasp91} velocity model through \texttt{obspy} \cite{obspy} and used to verify the association. 

The response of the detector to such events is modest. The energy of a distant teleseism reaching the site is concentrated below $\sim 1$~Hz, whereas the statistic is evaluated as a median over 2--15~Hz and responds only when the majority of the bins in that band rise together. Earthquakes of this class therefore produce no excursion of $z(t)$. The catalogue discussed in Section~\ref{sec:transients} is not appreciably contaminated by strong world-wide earthquakes but it is contaminated by the local and micro earthquakes of the region.

\section{Z-score detection}
\label{app:zscore}

\begin{figure}
    \includegraphics[width=\textwidth]{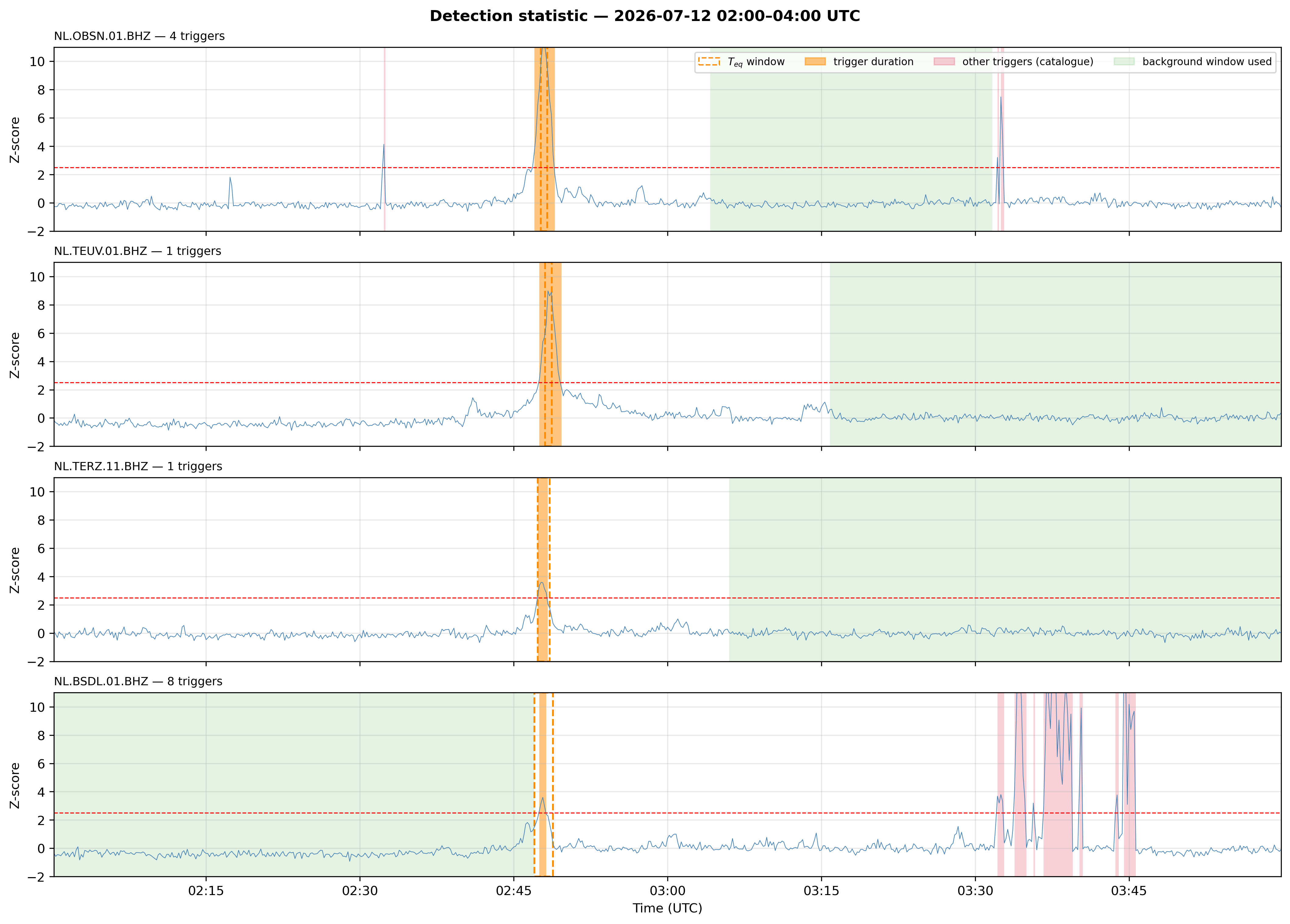}
\caption{Detection statistic $z(t)$ over the same chunk of Figure~\ref{fig:spectrograms_stacked}, for the borehole sensors. The dashed horizontal line is the detection threshold $z_{\rm thresh}=2.5$; the orange band is the trigger analyzed and the dashed orange lines delimit its $T_{\rm eq}$ window; the red bands are the other triggers of the same sensor, taken from the catalogue for the borehole and detected locally for the surface; the green band is the background window actually used, i.e.\ the longest stretch in which neither sensor of the station has a trigger, with a guard of 60\,s.}
\label{fig:detection_statistic_borehole}
\end{figure}

\begin{figure}
    \includegraphics[width=\textwidth]{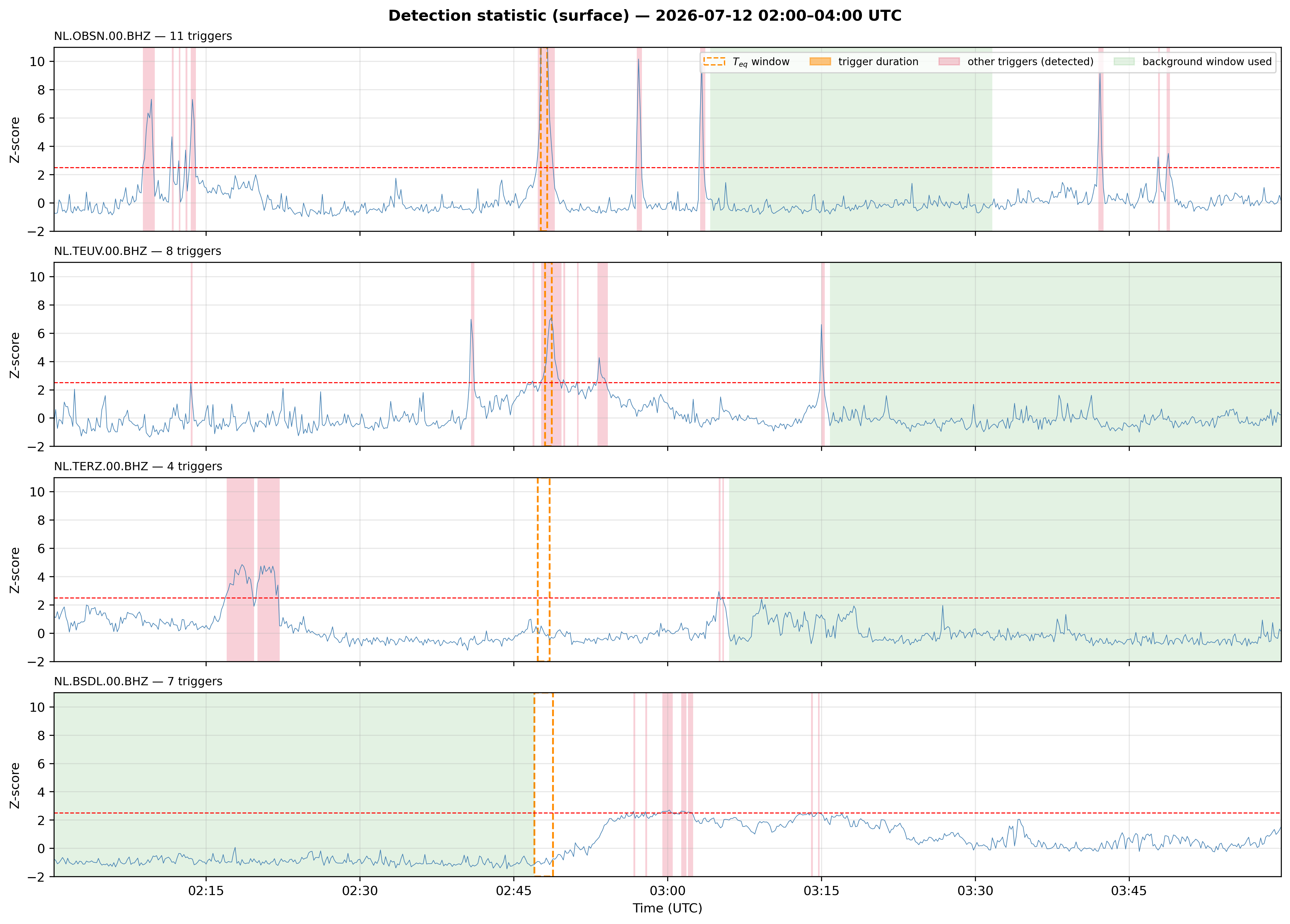}
\caption{Detection statistic $z(t)$ over the same chunk of Figure~\ref{fig:spectrograms_stacked}, for the co-located surface sensors. The dashed horizontal line is the detection threshold $z_{\rm thresh}=2.5$; the orange band is the trigger analyzed and the dashed orange lines delimit its $T_{\rm eq}$ window; the red bands are the other triggers of the same sensor, taken from the catalogue for the borehole and detected locally for the surface; the green band is the background window actually used, i.e.\ the longest stretch in which neither sensor of the station has a trigger, with a guard of 60\,s. At TERZ and BSDL the surface statistic does not cross the threshold inside the $T_{\rm eq}$ window of the borehole, and the event is treated there as an upper limit.}
\label{fig:detection_statistic_surface}
\end{figure}

Figures~\ref{fig:detection_statistic_borehole} and \ref{fig:detection_statistic_surface} show the detection statistic $z(t)$ over the chunk of Figure~\ref{fig:spectrograms_stacked}, for the borehole and for the co-located surface sensors respectively. The passage analyzed in Section~\ref{sec:single_event} is the excursion near 02:48 UTC, and the two figures make visible the quantities that enter the single-event analysis: the trigger returned by the detector, the $T_{\rm eq}$ window built around the envelope peak, the other triggers of the same sensor, and the background window actually used. The surface sensors carry between four and eleven triggers in the same two hours against one to eight at depth, which is what forces the background window to be defined on both sensors simultaneously. At TERZ and BSDL the surface statistic never crosses the threshold inside the $T_{\rm eq}$ window of the borehole, and the event is treated there as an upper limit.

The same figures illustrate the limits of the statistic. Because the normalisation is re-estimated on each block, the excursion of $z$ measures the excess over the local background and not an absolute level, so the height of a peak is not comparable between two stations, nor between day and night at the same station. The median across the band in \eqref{eq:score} responds only when more than half of the frequency bins rise together. The bottom panel of Figure~\ref{fig:detection_statistic_surface} shows a further limitation, of a different nature. Over the second half of the chunk the baseline of $z(t)$ at the BSDL surface sensor drifts upwards by about one unit without any trigger being involved. The ambient level of that sensor changes on a timescale comparable with the block itself, so the block does not contain a single stationary state but two.  The detector becomes less sensitive in the quiet stretch and closer to threshold in the noisy one, so that the same physical disturbance is recovered with different efficiency within a single block. 

The extent of a trigger is set by where the excursion crosses $z_{\rm thr}$ and therefore grows with the signal-to-noise ratio, which is why the trigger duration is not used as a measure of the transient and $T_{\rm eq}$ is defined on the band envelope instead. The equivalent duration is not free from the same effect. When the excess over the background is small, the residual envelope $\tilde{E}(t)$ of \eqref{eq:teq} is a weak signal on top of a fluctuating baseline, the subtraction of $E_0$ leaves a broad low-level pedestal inside the integration window, and $T_{\rm eq}$ is both inflated and considerably more dispersed than at high signal-to-noise ratio. The growth of the quoted $T_{\rm eq}$ from OBSN to BSDL in Figure~\ref{fig:event_bg_ratio}, for one and the same passage, is an illustration of this behaviour, and the larger median at TERZ in Figure~\ref{fig:trigger_statistics} should be read in the same way. At the opposite end, for the impulsive triggers of Section~\ref{sec:shorts} the measured $T_{\rm eq}$ is set by the 1~s smoothing of the envelope and by the 2--15~Hz band rather than by the signal, and it loses its meaning as a duration. The estimator is therefore reliable only in the intermediate regime of well-resolved, high signal-to-noise transients, which is the regime of the railway passages that dominate the catalogue.

\FloatBarrier

\printbibliography

\end{document}